\documentclass[aps,twocolumn]{revtex4}
\usepackage{amsmath}
\usepackage{epsfig}
\usepackage{multirow}
\usepackage{appendix}
\usepackage{amssymb}
\usepackage{epstopdf}

\begin{document}
	
\title{A coupled-channel quark model study of possible $\Xi_{cc}^{(*)} K^{(*)}$ molecular states}
\author{Ye Yan$^{1}$}
\author{Qi Huang$^2$}
\author{Yuheng Wu$^{3}$}
\author{Hongxia Huang$^2$}\email{hxhuang@njnu.edu.cn(Corresponding author)}
\author{Jialun Ping$^2$}

\affiliation{$^1$Department of Physics, Changzhou University of Information Technology, Changzhou 213164, China}
\affiliation{$^2$Department of Physics, Nanjing Normal University, Nanjing 210023, China}
\affiliation{$^3$Department of Physics, Yancheng Institute of Technology, Yancheng 224000, China}

\begin{abstract}
Inspired by the recent experimental discovery of doubly charmed baryons, we investigate the possible $\Xi_{cc}^{(*)}K^{(*)}$ molecular systems within the framework of the quark delocalization color screening model.
The energy spectra and scattering processes of the relevant baryon-meson systems are investigated to explore the dynamical properties of the possible molecular states.
The spectrum calculations predict three bound states, namely the $I(J^P)=0(1/2^{-})$ $\Xi_{cc}K$, the $I(J^P)=0(3/2^{-})$ $\Xi_{cc}^{*}K$, and the $I(J^P)=0(5/2^{-})$ $\Xi_{cc}^{*}K^{*}$ molecular states.
The scattering phase shift analysis further confirms two $\Xi_{cc}K^{*}$ resonance states with $I(J^P)=0(1/2^{-})$ and $0(3/2^{-})$, which originate from quasi-bound states through channel coupling.
In particular, the $I(J^P)=0(1/2^{-})$ $\Xi_{cc}K$ bound state is consistent with previous theoretical studies, making it one of the most promising candidates for future experimental searches.
\end{abstract}
	

\maketitle

\setcounter{totalnumber}{5}

\section{Introduction}

Understanding the nonperturbative dynamics of quantum chromodynamics (QCD) remains one of the central topics in hadron physics, where the spectroscopy of hadrons provides an important window into the strong interaction~\cite{Klempt:2009pi,Liu:2013waa,Richard:2016eis,Shepherd:2016dni,Guo:2017jvc,JPAC:2021rxu,Huang:2023jec,Liu:2024uxn}.
Owing to the presence of heavy charm quark, charmed hadrons provide a unique platform for investigating the interplay between heavy-quark dynamics and light-quark degrees of freedom, and have therefore attracted considerable theoretical and experimental attention~\cite{Olsen:2017bmm,Cheng:2015iom,Kato:2018ijx,Hosaka:2016pey,Chen:2016qju,Lebed:2016hpi,Liu:2019zoy,Brambilla:2019esw,Meng:2022ozq,Chen:2022asf}.

A major breakthrough was achieved in 2017, when the LHCb Collaboration reported the first observation of the doubly charmed baryon $\Xi_{cc}^{++}$ in the $\Lambda_c^+K^-\pi^+\pi^+$ final state~\cite{LHCb:2017iph}. 
The existence of the $\Xi_{cc}^{++}$ was subsequently confirmed in the $\Xi_c^+\pi^+$ decay channel~\cite{LHCb:2018pcs}.
In addition, extensive theoretical and experimental efforts have been devoted to exploring the spectroscopy, structure, production, and decay properties of doubly charmed baryons~\cite{LHCb:2019qed,LHCb:2019epo,LHCb:2018zpl,Ebert:2005xj,Roberts:2007ni,Lin:2011ti,Can:2013tna,Karliner:2014gca,Brown:2014ena,Lu:2017meb,Xiao:2017udy}.
Very recently, the LHCb Collaboration reported the first observation of the isospin partner, $\Xi_{cc}^{+}$, in the $\Lambda_c^+K^-\pi^+$ final state~\cite{LHCb:2026pxn}. 
The observation of both members of the doubly charmed isodoublet marks an important milestone in doubly charmed baryon spectroscopy and further stimulates the exploration of exotic hadrons containing doubly charmed baryons.

Besides the remarkable progress in doubly charmed baryons, extensive experimental evidence has also been accumulated for exotic multiquark states. 
In the tetraquark sector, a number of candidates involving charm quarks have been reported in recent years, including the $T_{cc}^{+}(3875)$, $D_{s0}^{*}(2317)$, $D_{s1}(2460)$, and $T_{c\bar{s}}^{a0(++)}(2900)$ states~\cite{BaBar:2003oey,CLEO:2003ggt,LHCb:2020bls,LHCb:2020pxc,LHCb:2021vvq,LHCb:2022sfr,LHCb:2022lzp}.
Their underlying structures have attracted considerable attention, and have been interpreted as hadronic molecules, compact tetraquarks, conventional quark-antiquark mesons, or mixtures of these configurations in different theoretical approaches.

Among them, the $T_{cc}^{+}(3875)$ state, located extremely close to the $D D^*$ threshold, is widely regarded as one of the most promising hadronic molecular candidates~\cite{Du:2021zzh,Albaladejo:2021vln,Meng:2021jnw,Feijoo:2021ppq,Fleming:2021wmk,Ling:2021bir,Chen:2021vhg,Wu:2021kbu,Padmanath:2022cvl,Lyu:2023xro,Wang:2021mma,Dong:2021bvy,Dai:2023kwv}.
Similarly, the $D^*_{s0}(2317)$ and $D_{s1}(2460)$, lying below the $D K$ and $D^* K$ thresholds, respectively, have long been suggested to contain sizable hadronic molecular components~\cite{Barnes:2003dj,Kolomeitsev:2003ac,vanBeveren:2003kd,Guo:2006fu,Gamermann:2006nm,Faessler:2007gv,Cleven:2014oka,Mohler:2013rwa,Lang:2014yfa,Ortega:2016mms,Du:2017ttu}.
Recently, the LHCb Collaboration observed the isovector $T_{c\bar{s}0}^{*}(2900)$ states, including the doubly charged $T_{c\bar{s}0}^{*}(2900)^{++}$ and its neutral partner $T_{c\bar{s}0}^{*}(2900)^{0}$, in the $D_s^+\pi^\pm$ invariant mass distributions from $B$ meson decays~\cite{LHCb:2022sfr,LHCb:2022lzp}. 
The measured quantum number $J^P=0^+$ and the proximity of their masses to the $D^*K^*$ threshold have motivated various interpretations, including $D^*K^*$ molecular states and compact tetraquark $c\bar{s}q\bar{q}$ configurations~\cite{Chen:2022svh,Agaev:2022eyk,Yue:2022mnf,Duan:2023lcj,Ke:2022ocs,Liu:2022hbk,Yang:2023evp,Lian:2023cgs,Wei:2022wtr,Ortega:2023azl,Molina:2022jcd,Duan:2023qsg,Huang:2023fvj,Wang:2023hpp}.

In addition to the open-charm tetraquark candidates discussed above, exotic states containing anticharm quarks have also attracted extensive attention. 
The $X_0(2900)$ and $X_1(2900)$ states, now denoted as $T_{\bar{c}\bar{s}0}^{*}(2870)^{0}$ and $T_{\bar{c}\bar{s}1}^{*}(2900)^{0}$ under the new naming convention~\cite{ParticleDataGroup:2026aaa}, were observed by the LHCb Collaboration in the $D^-K^+$ invariant mass spectrum from the decays $B^+\rightarrow D^+D^-K^+$~\cite{LHCb:2020pxc,LHCb:2020bls}. 
These states possess the minimal quark content $ud\bar{c}\bar{s}$ and have provided important candidates for studying open anti-charm tetraquark systems. 
Their masses, decay properties, production mechanisms, and possible internal structures have been extensively investigated in various theoretical approaches~\cite{Molina:2010tx,Agaev:2020nrc,Albuquerque:2020ugi,Burns:2020epm,Chen:2020aos,Chen:2020eyu,He:2020btl,He:2020jna,Hu:2020mxp,Huang:2020ptc,Karliner:2020vsi,Liu:2020nil,Liu:2020orv,Lu:2020qmp,Tan:2020cpu,Wang:2020prk,Wang:2020xyc,Xiao:2020ltm,Xue:2020vtq,Zhang:2020oze,Chen:2021tad,Wang:2021lwy,Bayar:2022wbx,Dai:2022htx,Lin:2022eau}.
For instance, the $X_0(2900)$ state can be interpreted as an $S$-wave $\bar{D}^{*}K^{*}$ molecular state~\cite{Agaev:2020nrc,He:2020btl,Xue:2020vtq,Chen:2020aos,Liu:2020nil,Hu:2020mxp,Wang:2021lwy} or a $ud\bar{c}\bar{s}$ state~\cite{Wang:2020xyc,Albuquerque:2020ugi,Lu:2020qmp,Wang:2020prk,Zhang:2020oze,Tan:2020cpu} in many theoretical studies, whereas the internal structure of the $X_1(2900)$ state remains more controversial, with various interpretations including a $P$-wave $\bar{D}^{*}K^{*}$ molecular state~\cite{Wang:2021lwy}, a $\bar{D}_{1}K^{*}$ molecular configuration~\cite{He:2020btl}, a $P$-wave compact $\bar{c}\bar{s}ud$ tetraquark state~\cite{Chen:2020aos}, and other possible interpretations~\cite{Burns:2020epm,Liu:2020orv}.

The experimental discovery of hidden-charm pentaquark candidates has further enriched the multiquark spectrum. 
The LHCb Collaboration first observed the $P_c$ states in the $J/\psi p$ invariant mass spectrum from the decay $\Lambda_b^0\rightarrow J/\psi pK^-$~\cite{LHCb:2015yax}, and later reported the additional $P_c(4312)$, $P_c(4440)$, and $P_c(4457)$ structures with larger data samples~\cite{LHCb:2019kea}. 
More recently, the strange hidden-charm pentaquark candidate $P_{cs}(4338)$ was observed in the $J/\psi\Lambda$ invariant mass spectrum~\cite{LHCb:2022ogu}. 
These discoveries have demonstrated the rich dynamics of pentaquark systems and motivated the exploration of new types of exotic configurations.
Motivated by the recent progress in the study of tetraquark states, the doubly charmed pentaquark systems with the quark content $ccqq\bar{s}$ can be naturally constructed. 
Owing to heavy-antiquark--diquark symmetry (HADS), the doubly charmed diquark $(cc)$ and the heavy anticharm quark $\bar c$ play analogous roles in their interactions with the light degrees of freedom~\cite{Savage:1990di,Hu:2005gf,Guo:2013sya}. 
Therefore, the open anti-charm tetraquark systems may have doubly charmed partners through the replacement:
\begin{align}
	T_{\bar{c}\bar{s}(0/1)}^{*}(qq\bar{c}\bar{s})
	\xrightarrow{\bar{c}\rightarrow cc}
	\Xi_{cc}^{(*)}K^{(*)}(ccqq\bar{s}).
\end{align}
From the hadronic perspective, the doubly charmed pentaquark systems can also be viewed as the baryonic counterparts of the related possible charm-strange molecular candidates:
\begin{equation}
	\begin{pmatrix}
		D_{s0}^{*}~(DK)\\
		D_{s1}~(D^{*}K)\\
		T_{c\bar{s}0}^{*}~(D^{*}K^{*})
	\end{pmatrix}
	\xrightarrow{D^{(*)}\rightarrow\Xi_{cc}^{(*)}}
	\Xi_{cc}^{(*)}K^{(*)}.
\end{equation}
Therefore, the $\Xi_{cc}^{(*)}K^{(*)}$ systems with the exotic quark content $ccqq\bar{s}$ provide a natural extension of the recently discovered multiquark states and offer an important platform for exploring doubly charmed hadronic molecules.

Possible pentaquark systems with the quark content $qqcc\bar{s}$ have attracted increasing theoretical attention.
The existence of a $\Xi_{cc}K$ molecular state was first predicted within a coupled-channel approach~\cite{Hofmann:2005sw}. 
Subsequently, a bound $\Xi_{cc}K$ state was also obtained within chiral effective field theory~\cite{Guo:2017vcf}. 
Within the one-boson-exchange model, four $\Xi_{cc}^{(*)}K^{(*)}$ molecular candidates were predicted, and their electromagnetic properties were further investigated~\cite{Sheng:2024hkf}. 
More recently, the extended local hidden gauge approach also predicted four $\Xi_{cc}^{(*)}K^{(*)}$ molecular states, although the spin-parity quantum numbers of most states remain degenerate owing to the symmetry of the framework~\cite{Wang:2025hhx}. 
Besides these molecular studies, compact $ccqq\bar{s}$ pentaquark states have also been investigated within the color-magnetic interaction (CMI) model, SU(3) symmetry analysis, and mass splitting model, where rich spectra of doubly charmed pentaquark candidates were predicted~\cite{Zhou:2018bkn,Xing:2021yid,Li:2025omw,Rostami:2026jyz}. 
These studies consistently indicate that the $\Xi_{cc}K$ system is one of the most promising doubly charmed molecular candidates. 
However, the dynamical properties of this system still deserve further investigation at the quark level.

To explore the possible doubly charmed molecular states with the quark content $ccqq\bar{s}$, we employ the quark delocalization color screening model (QDCSM), which was originally proposed to explain the similarities between nuclear and molecular forces~\cite{Wu:1996fm}. 
The QDCSM has been successfully applied to describe the baryon-baryon interactions, the properties of the deuteron, and various exotic hadronic systems~\cite{Ping:1998si,Ping:2000dx,Wu:1998wu,Pang:2001xx,Huang:2023jec,Yan:2022nxp,Yan:2023tvl}. 
In particular, the model predicted a series of hidden-charm molecular pentaquark states associated with the $P_c$ structures observed by the LHCb Collaboration~\cite{Huang:2015uda,Huang:2018wed}, and has also been employed to investigate open-charm multiquark systems, where the $X_0(2900)$ can be interpreted as a $D^{*}K^{*}$ molecular state within the QDCSM framework~\cite{Xue:2020vtq}. 
These successful applications motivate us to extend the QDCSM to the $ccqq\bar{s}$ systems in search of possible doubly charmed molecular states.

The remainder of this paper is organized as follows. 
Section~\ref{2} briefly introduces the QDCSM. 
Section~\ref{3} presents the adiabatic potential, energy-spectrum, and scattering phase shift analyses of the baryon-meson molecular $ccqq\bar{s}$ systems.
Finally, the conclusions are given in Sec.~\ref{4}.

\section{THEORETICAL FORMALISM}
\label{2}

\subsection{Quark delocalization color screening model}
\label{21}

In this section, we briefly introduce the salient features of the used quark model.
The QDCSM is an extension of the quark cluster model~\cite{Wang:1992wi} and the general form of the pentaquark Hamiltonian is given by:
	\begin{align}
	H=&\sum_{i=1}^5\left(m_i+\frac{\boldsymbol{p}_{i}^{2}}{2m_i}\right)-T_{\text{c.m.}} +\sum_{j>i=1}^5 V(\boldsymbol{r}_{ij}),
\end{align}
where $m_i$ is the quark mass, $\boldsymbol{p}_{i}$ is the momentum of the quark, and $T_{\text{c.m.}}$ is the center-of-mass kinetic energy.
The dynamics of the hexaquark system is driven by two-body potentials, including color confinement ($V_{\mathrm{conf}}$), perturbative one-gluon exchange interaction ($V_{\mathrm{OGE}}$), and dynamical chiral symmetry breaking ($V_{\chi}$).
\begin{align}
	V(\boldsymbol{r}_{ij})= & V_{\mathrm{conf}}(\boldsymbol{r}_{ij})+V_{\mathrm{OGE}}(\boldsymbol{r}_{ij})+V_{\chi}(\boldsymbol{r}_{ij}).
\end{align}

Here, a phenomenological color screening confinement potential ($V_{\mathrm{conf}}$) is used as:
\begin{align}
	V_{\mathrm{conf}}(\boldsymbol{r}_{ij}) = & -a_{c}\boldsymbol{\lambda}_{i}^{c} \cdot \boldsymbol{\lambda}_{j}^{c}\left[  f(\boldsymbol{r}_{ij})+V_{0}\right],
\end{align}
\begin{align}
	f(\boldsymbol{r}_{ij}) =& \left\{\begin{array}{l}
		\boldsymbol{r}_{i j}^{2}, ~~~~~~~~~~~~~ ~i,j ~\text {occur in the same cluster } \\
		\frac{1-\text{e}^{-\mu_{q_{i}q_{j}} \boldsymbol{r}_{i j}^{2}}}{\mu_{q_{i}q_{j}}},  ~~~i,j ~\text {occur in different cluster }
	\end{array}\right.   \nonumber
\end{align}
where $a_c$, $V_{0}$ and $\mu_{q_{i}q_{j}}$ are model parameters, and $\boldsymbol{\lambda}^{c}$ stands for the SU(3) color Gell-Mann matrices.
Among them, the color screening parameter $\mu_{q_{i}q_{j}}$ is determined by fitting the deuteron properties, nucleon-nucleon scattering phase shifts, and hyperon-nucleon scattering phase shifts, respectively, with $\mu_{qq}=0.45$, $\mu_{qs}=0.19$, and $\mu_{ss}=0.08~$fm$^{-2}$, satisfying the relation $\mu_{qs}^{2}=\mu_{qq}\mu_{ss}$~\cite{Chen:2011zzb}.

In the present work, we mainly focus on the low-lying negative parity system of the $S$-wave, so the spin-orbit and tensor interactions are not included.
The one-gluon exchange potential ($V_{\mathrm{OGE}}$), which includes Coulomb and chromomagnetic interactions, is written as:
\begin{align}
	V_{\mathrm{OGE}}(\boldsymbol{r}_{ij})=&\frac{1}{4}\alpha_{s} \boldsymbol{\lambda}_{i}^{c} \cdot \boldsymbol{\lambda}_{j}^{c}  \\
	&\times \left[\frac{1}{r_{i j}}-\frac{\pi}{2} \delta\left(\mathbf{r}_{i j}\right)\left(\frac{1}{m_{i}^{2}}+\frac{1}{m_{j}^{2}}+\frac{4 \boldsymbol{\sigma}_{i} \cdot \boldsymbol{\sigma}_{j}}{3 m_{i} m_{j}}\right)\right],   \nonumber \label{Voge}
\end{align}
where $\boldsymbol{\sigma}$ is the Pauli matrices and $\alpha_{s}$ is the quark-gluon coupling constant.

The dynamical breaking of chiral symmetry results in the SU(3) Goldstone boson exchange interactions appear between constituent light quarks $u, d$, and $s$.
Hence, the chiral interaction is expressed as:
\begin{align}
	V_{\chi}(\boldsymbol{r}_{ij})= & V_{\pi}(\boldsymbol{r}_{ij})+V_{K}(\boldsymbol{r}_{ij})+V_{\eta}(\boldsymbol{r}_{ij}).
\end{align}
Among them:
\begin{align}
	V_{\pi}\left(\boldsymbol{r}_{i j}\right) =&\frac{g_\text{ch}^{2}}{4 \pi} \frac{m_{\pi}^{2}}{12 m_{i} m_{j}} \frac{\Lambda_{\pi}^{2}}{\Lambda_{\pi}^{2}-m_{\pi}^{2}} m_{\pi}\left[Y\left(m_{\pi} \boldsymbol{r}_{i j}\right)\right. \nonumber \\
	&\left.-\frac{\Lambda_{\pi}^{3}}{m_{\pi}^{3}} Y\left(\Lambda_{\pi} \boldsymbol{r}_{i j}\right)\right]\left(\boldsymbol{\sigma}_{i} \cdot \boldsymbol{\sigma}_{j}\right) \sum_{a=1}^{3}\left(\boldsymbol{\lambda}_{i}^{a} \cdot \boldsymbol{\lambda}_{j}^{a}\right),
\end{align}
\begin{align}
	V_{K}\left(\boldsymbol{r}_{i j}\right) =&\frac{g_\text{ch}^{2}}{4 \pi} \frac{m_{K}^{2}}{12 m_{i} m_{j}} \frac{\Lambda_{K}^{2}}{\Lambda_{K}^{2}-m_{K}^{2}} m_{K}\left[Y\left(m_{K} \boldsymbol{r}_{i j}\right)\right. \nonumber \\
	&\left.-\frac{\Lambda_{K}^{3}}{m_{K}^{3}} Y\left(\Lambda_{K} \boldsymbol{r}_{i j}\right)\right]\left(\boldsymbol{\sigma}_{i} \cdot \boldsymbol{\sigma}_{j}\right) \sum_{a=4}^{7}\left(\boldsymbol{\lambda}_{i}^{a} \cdot \boldsymbol{\lambda}_{j}^{a}\right),
\end{align}
\begin{align}
	V_{\eta}\left(\boldsymbol{r}_{i j}\right) =&\frac{g_\text{ch}^{2}}{4 \pi} \frac{m_{\eta}^{2}}{12 m_{i} m_{j}} \frac{\Lambda_{\eta}^{2}}{\Lambda_{\eta}^{2}-m_{\eta}^{2}} m_{\eta}\left[Y\left(m_{\eta} \boldsymbol{r}_{i j}\right)\right. \nonumber \\
	&\left.-\frac{\Lambda_{\eta}^{3}}{m_{\eta}^{3}} Y\left(\Lambda_{\eta} \boldsymbol{r}_{i j}\right)\right]\left(\boldsymbol{\sigma}_{i} \cdot \boldsymbol{\sigma}_{j}\right)\left[\cos \theta_\text{p}\left(\boldsymbol{\lambda}_{i}^{8} \cdot \boldsymbol{\lambda}_{j}^{8}\right)\right.  \nonumber \\
	&\left.-\sin \theta_\text{p}\left(\boldsymbol{\lambda}_{i}^{0} \cdot \boldsymbol{\lambda}_{j}^{0}\right)\right],
\end{align}
where $Y(x) = \text{e}^{-x}/x$ is the standard Yukawa function.
The physical $\eta$ meson is considered by introducing the angle $\theta_\text{p}$ instead of the octet one.
 The $\boldsymbol{\lambda}^a$ are the SU(3) flavor Gell-Mann matrices.
The values of $m_\pi$, $m_K$, and $m_\eta$ are the masses of the SU(3) Goldstone bosons, which adopt the experimental values~\cite{ParticleDataGroup:2026aaa}.

In the present work, the same chiral coupling constant $g_\text{ch}$ as in the light-quark sector is adopted for consistency, which is determined from the $\pi N N$ coupling constant through~\cite{Vijande:2004he}:
\begin{align}
	\frac{g_\text{ch}^{2}}{4 \pi} & = \left(\frac{3}{5}\right)^{2} \frac{g_{\pi N N}^{2}}{4 \pi} \frac{m_{u, d}^{2}}{m_{N}^{2}}.
\end{align}
The other symbols in the above expressions have their usual meanings.
The parameters used in this work are identical to those adopted in our previous study of excited $\Xi$ states from a pentaquark perspective~\cite{Yan:2024usf}, allowing for a consistent treatment of strange hadronic systems.
Table~\ref{parameters} lists the model parameters, while Table~\ref{hadrons} summarizes the calculated masses of the hadrons.

\begin{table}[htb]
	\caption{\label{parameters}Model parameters used in this work:
		$m_{\pi} = 0.7$, $m_{K} = 2.51$, $m_{\eta} = 2.77$, $\Lambda_{\pi} = 4.2$, $\Lambda_{K} = 5.2$, $\Lambda_{\eta} = 5.2$ fm$^{-1}$, $g_\text{ch}^2/(4\pi)$ = 0.54.}
	\begin{tabular}{cccccc} \hline\hline
		~~~~$b$~~~~ & ~~$m_{q}$~~ & ~~~$m_{s}$~~~  & ~~~$V_{0_{qq}}$~~~~&~~~$V_{0_{q\bar{q}}}$~~~~& ~~~$ a_c$~~~   \\
		(fm)        & (MeV)       & (MeV)          & (fm$^{-2}$)        & (fm$^{-2}$)             & ~(MeV\,fm$^{-2}$)~ \\
		0.518       & 313         & 573            &  $-$1.288          & $-$0.201                &  58.03 \\ \hline
		$\alpha_{s_{qq}}$&$\alpha_{s_{qc}}$& $\alpha_{s_{cc}}$ &$\alpha_{s_{q\bar{s}}} $ & $\alpha_{s_{c\bar{s}}}$ & \\
		0.565            & 0.467           &  0.213            &    1.783                  & 1.513                 & \\ \hline\hline	
	\end{tabular}
\end{table}

\begin{table}[htb]
	\caption{The masses (in MeV) of the baryons and mesons. Experimental values are taken from the Particle Data Group~\cite{ParticleDataGroup:2026aaa}.}
	\begin{tabular}{c c c |c c c}
		\hline \hline
		~Baryon~ & ~$M^{\text{Exp}}$~ & ~$M^{\text{Theo}}$~ & ~Meson~ & ~$M^{\text{Exp}}$~ & ~$M^{\text{Theo}}$~ \\ \hline
		$N$          & 939   & 939  & $\pi$ & 139 & 139 \\
		$\Delta$     & 1232  & 1232 & $\rho$ & 770  & 890  \\
		$\Lambda$    & 1115  & 1123 & $K$ & 495  & 495  \\
		$\Sigma$     & 1189  & 1238 & $K^*$ & 892 & 892 \\
		$\Sigma^*$   & 1385  & 1360 & $\eta^\prime$ & 958 & 851 \\
		$\Lambda_c$  & 2286  & 2286 & $\phi$ & 1020 & 1020 \\
		$\Sigma_c$   & 2455  & 2464 & $D$ & 1869 & 1869 \\  
		$\Sigma_c^*$ & 2520  & 2489 & $D^*$ & 2007 & 1983 \\  
		$\Xi_{cc}$   & 3621  & 3766 & $D_s$ & 1968 & 2053     \\
		$\Xi_{cc}^*$ & ---   & 3791 & $D_s^*$ & 2112 & 2112          \\
		\hline\hline
	\end{tabular}
	\label{hadrons}
\end{table}

In the QDCSM, quark delocalization was introduced to enlarge the model variational space to take into account the mutual distortion or the internal excitations of nucleons in the course of interaction.
It is realized by specifying the single-particle orbital wave function of the QDCSM as a linear combination of left and right Gaussians, the single-particle orbital wave functions used in the ordinary quark cluster model
\begin{eqnarray}
	\psi_{\alpha}(\boldsymbol {S_{i}} ,\epsilon) & = & \left(
	\phi_{\alpha}(\boldsymbol {S_{i}})
	+ \epsilon \phi_{\alpha}(-\boldsymbol {S_{i}})\right) /N(\epsilon), \nonumber \\
	\psi_{\beta}(-\boldsymbol {S_{i}} ,\epsilon) & = &
	\left(\phi_{\beta}(-\boldsymbol {S_{i}})
	+ \epsilon \phi_{\beta}(\boldsymbol {S_{i}})\right) /N(\epsilon), \nonumber \\
	N(S_{i},\epsilon) & = & \sqrt{1+\epsilon^2+2\epsilon \text{e}^{-S_i^2/4b^2}}. \label{1q}
\end{eqnarray}
It is worth noting that the mixing parameter $\epsilon$ is not an adjusted one but determined variationally by the dynamics of the multiquark system itself.
In this way, the multiquark system chooses its favorable configuration in the interacting process.
This mechanism has been used to explain the crossover transition between the hadron phase and quark-gluon plasma phase~\cite{Xu:2007oam}.

\subsection{Resonating group method for bound-state and scattering process}
\label{22}
The resonating group method (RGM)~\cite{Wheeler:1937zza,Kamimura:1977okl} and generating coordinates method~\cite{Hill:1952jb,Griffin:1957zza} are used to carry out a dynamical calculation.
The main feature of the RGM for two-cluster systems is that it assumes that two clusters are frozen inside, and only considers the relative motion between the two clusters.
So the conventional ansatz for the two-cluster wave functions is
\begin{equation}
	\psi_{5q} = {\cal A }\left[[\phi_{B}\phi_{M}]^{[\sigma]IS}\otimes\chi(\boldsymbol{R})\right]^{J}, \label{5q}
\end{equation}
where the symbol ${\cal A }$ is the antisymmetrization operator, and ${\cal A} = 1-P_{14}-P_{24}-P_{34}$. $[\sigma]=[222]$ gives the total color symmetry and all other symbols have their usual meanings.
$\phi_{B}$ and $\phi_{M}$ are the $q^{3}$ and $\bar{q}q$ cluster wave functions, respectively.
From the variational principle, after variation with respect to the relative motion wave function $\chi(\boldsymbol{R})=\sum_{L}\chi_{L}(\boldsymbol{R})$, one obtains the RGM equation:
\begin{equation}
	\int H(\boldsymbol{R},\boldsymbol{R}')\chi(\boldsymbol{R'}) \, \text{d} \boldsymbol{R'}=E\int N(\boldsymbol{R},\boldsymbol{R}')\chi(\boldsymbol{R}') \, \text{d} \boldsymbol{R'},  \label{RGM eq}
\end{equation}
where $H(\boldsymbol{R},\boldsymbol{R}')$ and $N(\boldsymbol{R},\boldsymbol{R}')$ are Hamiltonian and norm kernels.
By solving the RGM equation, we can get the energies $E$ and the wave functions.
In fact, it is not convenient to work with the RGM expressions.
Then, we expand the relative motion wave function $\chi(\boldsymbol{R})$ by using a set of Gaussians with different centers
\begin{align}
	\chi(\boldsymbol{R}) =& \frac{1}{\sqrt{4 \pi}}\left(\frac{6}{5 \pi b^{2}}\right)^{3/4} \sum_{i,L,M} C_{i,L} \nonumber     \\
	&\times \int \exp \left[-\frac{3}{5 b^{2}}\left(\boldsymbol{R}-\boldsymbol{S}_{i}\right)^{2}\right] Y_{L,M}\left(\hat{\boldsymbol{S}}_{i}\right) \, \text{d} \Omega_{\boldsymbol{S}_{i}}
\end{align}
where $L$ is the orbital angular momentum between two clusters, and $\boldsymbol {S_{i}}$, $i=1,2,...,n$ are the generator coordinates, which are introduced to expand the relative motion wave function. By including the center-of-mass motion:
\begin{equation}
	\phi_{C} (\boldsymbol{R}_{C}) = (\frac{5}{\pi b^{2}})^{3/4} \text{e}^{-\tfrac{5\boldsymbol{R}^{2}_{C}}{2b^{2}}},
\end{equation}
the ansatz Eq.~(\ref{5q}) can be rewritten as
\begin{align}
	\psi_{5 q} =& \mathcal{A} \sum_{i,L} C_{i,L} \int \frac{\text{d} \Omega_{\boldsymbol{S}_{i}}}{\sqrt{4 \pi}} \prod_{\alpha=1}^{3} \phi_{\alpha}\left(\boldsymbol{S}_{i}\right) \prod_{\beta=4}^{5} \phi_{\beta}\left(-\boldsymbol{S}_{i}\right) \nonumber \\
	& \times \left[\left[\chi_{I_{1} S_{1}}\left(B\right) \chi_{I_{2} S_{2}}\left(M\right)\right]^{I S} Y_{LM}\left(\hat{\boldsymbol{S}}_{i}\right)\right]^{J} \nonumber \\
	& \times \left[\chi_{c}\left(B\right) \chi_{c}\left(M\right)\right]^{[\sigma]}, \label{5q2}
\end{align}
where $\chi_{I_{1}S_{1}}$ and $\chi_{I_{2}S_{2}}$ are the product of the flavor and spin wave functions, and $\chi_{c}$ is the color wave function.
These will be shown in detail later.
$\phi_{\alpha}(\boldsymbol{S}_{i})$ and $\phi_{\beta}(-\boldsymbol{S}_{i})$ are the single-particle orbital wave functions with different reference centers,
\begin{align}
	\phi_{\alpha}\left(\boldsymbol{S}_{i}\right) & = \left(\frac{1}{\pi b^{2}}\right)^{3 / 4} e^{-\tfrac{1}{2 b^{2}}\left(r_{\alpha}-\tfrac{2}{5} \boldsymbol{S}_{i}\right)^{2}}, \nonumber \\
	\phi_{\beta}\left(\boldsymbol{-S}_{i}\right) & = \left(\frac{1}{\pi b^{2}}\right)^{3 / 4} e^{-\tfrac{1}{2 b^{2}}\left(r_{\beta}+\tfrac{3}{5} \boldsymbol{S}_{i}\right)^{2}}.
\end{align}
With the reformulated ansatz Eq.~(\ref{5q2}), the RGM Eq.~(\ref{RGM eq}) becomes an algebraic eigenvalue equation:
\begin{equation}
	\sum_{j} C_{j}H_{i,j}= E \sum_{j} C_{j}N_{i,j},
\end{equation}
where $H_{i,j}$ and $N_{i,j}$ are the Hamiltonian matrix elements and overlaps, respectively.
By solving the generalized eigenproblem, we can obtain the energy and the corresponding wave functions of the pentaquark systems.

For a scattering problem, the relative wave function is expanded as
\begin{align}
	\chi_{L}(\mathbf{R}) & =\sum_{i} C_{i} \frac{\tilde{u}_{L}\left(\boldsymbol{R}, \boldsymbol{S}_{i}\right)}{\boldsymbol{R}} Y_{L,M}(\hat{\boldsymbol{R}}),
\end{align}
with
\begin{align}
	\tilde{u}_{L}\left(\boldsymbol{R}, \boldsymbol{S}_{i}\right) & = \left\{\begin{array}{ll}
		\alpha_{i} u_{L}\left(\boldsymbol{R}, \boldsymbol{S}_{i}\right), & \boldsymbol{R} \leq \boldsymbol{R}_{C} \\
		{\left[h_{L}^{-}(\boldsymbol{k}, \boldsymbol{R})-s_{i} h_{L}^{+}(\boldsymbol{k}, \boldsymbol{R})\right] R_{A B},} & \boldsymbol{R} \geq \boldsymbol{R}_{C}
	\end{array}\right.
\end{align}
where
\begin{align}
	u_{L}\left(\boldsymbol{R}, \boldsymbol{S}_{i}\right)= & \sqrt{4 \pi}\left(\frac{6}{5 \pi b^{2}}\right)^{3 / 4} \boldsymbol{R} \, \text{e}^{-\tfrac{3}{5 b^{2}}\left(\boldsymbol{R}-\boldsymbol{S}_{i}\right)^{2}} \nonumber \\
	& \times \mathrm{i}^{L} j_{L}\left(-\mathrm{i} \frac{6}{5 b^{2}} S_{i}\right).
\end{align}

$h^{\pm}_L$ are the $L$th spherical Hankel functions, $k$ is the momentum of the relative motion with $k=\sqrt{2 \mu E_\text{ie}}$, $\mu$ is the reduced mass of two hadrons of the open channel, $E_\text{ie}$ is the incident energy of the relevant open channels, which can be written as $E_\text{ie} = E_\text{total} - E_\text{th}$, where $E_\text{total}$ denotes the total energy, and $E_\text{th}$ represents the threshold of the open channel.
$R_C$ is a cutoff radius beyond which all the strong interaction can be disregarded.
Additionally, $\alpha_i$ and $s_i$ are complex parameters that are determined by the smoothness condition at $R = R_C$ and $C_i$ satisfy $\sum_i C_i = 1$. After performing the variational procedure, a $L$th partial-wave equation for the scattering problem can be deduced as
\begin{align}
	\sum_j \mathcal{L}_{i j}^L C_j &= \mathcal{M}_i^L(i=0,1, \ldots, n-1), \label{sp}
\end{align}
with
\begin{align}
	\mathcal{L}_{i j}^L&=\mathcal{K}_{i j}^L-\mathcal{K}_{i 0}^L-\mathcal{K}_{0 j}^L+\mathcal{K}_{00}^L, \nonumber \\
	\mathcal{M}_i^L&=\mathcal{K}_{00}^L-\mathcal{K}_{i 0}^L,
\end{align}
and
\begin{align}
	\mathcal{K}_{i j}^L= & \left\langle\hat{\phi}_A \hat{\phi}_B \frac{\tilde{u}_L\left(\boldsymbol{R}^{\prime}, \boldsymbol{S}_i\right)}{\boldsymbol{R}^{\prime}} Y_{L,M}\left(\boldsymbol{R}^{\prime}\right)|H-E|\right. \nonumber \\
	& \left.\times \mathcal{A}\left[\hat{\phi}_A \hat{\phi}_B \frac{\tilde{u}_L\left(\boldsymbol{R}, \boldsymbol{S}_j\right)}{\boldsymbol{R}} Y_{L,M}(\boldsymbol{R})\right]\right\rangle .
\end{align}
By solving Eq.~(\ref{sp}), we can obtain the expansion coefficients $C_i$, then the $S$-matrix element $S_L$ and the phase shifts $\delta_L$ are given by
\begin{align}
	S_L = \mathrm{e}^{2 \mathrm{i} \delta_L} = \sum_i C_i s_i.
\end{align}

Resonances are unstable particles usually observed as bell-shaped structures in scattering cross sections of their open channels.
For a simple narrow resonance, its fundamental properties correspond to the visible cross section features: mass $M$ is at the peak position, and decay width $\Gamma$ is the half-width of the bell shape.
The cross section $\sigma_{L}$ and the scattering phase shifts $\delta_{L}$ have relations
\begin{align}
	\sigma_L&=\frac{4 \pi}{k^2}(2 L+1) \sin ^2 \delta_L.
\end{align}
Therefore, resonances can also usually be observed in the scattering phase shift, where the phase shift of the scattering channels rises through $\pi/2$ at a resonance mass.
We can obtain a resonance mass at the position of the phase shift of $\pi/2$.
The decay width is the mass difference between the phase shift of $3\pi/4$ and $\pi/4$.

\section{Results and discussions}
\label{3}

we perform a comprehensive investigation of the hadronic molecular states with the quark contents $ccqq\bar{s}$ within the framework of the QDCSM. 
The five-quark dynamics are treated using the RGM, while the scattering observables are further analyzed with the Kohn-Hulthén-Kato (KHK) method. 
To investigate the interaction properties of the system, we systematically calculate the adiabatic potentials, the energy spectra, and the scattering phase shifts, aiming to identify possible bound and resonance states.

The present analysis is carried out in the hadronic molecular picture by considering all relevant baryon-meson channels allowed by the quantum numbers, including $\Xi_{cc}^{(*)}K^{(*)}$, $\Lambda_{c}D_{s}^{(*)}$, and $\Sigma_{c}^{(*)}D_{s}^{(*)}$. 
We focus on the lowest $S$-wave configurations with negative parity. 
Accordingly, the total angular momentum $J$ coincides with the total spin of the system, leading to the possible quantum numbers $J^{P}=1/2^{-}$, $3/2^{-}$, and $5/2^{-}$. 
Both isospin sectors, $I=0$ and $I=1$, are investigated systematically.

In the following, we first present the adiabatic potentials to reveal the interaction features of the various baryon-meson channels. 
We then discuss the results of the single-channel and coupled-channel spectrum calculations, followed by an analysis of the scattering phase shifts to further examine the resonance properties of the $ccqq\bar{s}$ system.

\subsection{Adiabatic potentials}
\label{31}

Before performing dynamical calculations, we first investigate the adiabatic potentials to obtain a qualitative understanding of the interactions between the baryon and meson clusters.
The adiabatic potential reflects the interaction between two hadronic clusters as a function of their separation and provides useful insight into whether a molecular configuration is dynamically favored. 
In general, an attractive potential well is a prerequisite for the formation of molecular bound states and may also favor the emergence of resonance states, whereas a purely repulsive interaction usually disfavors such structures.

The adiabatic potential is defined as the difference between the adiabatic energy and its asymptotic value,
\begin{align}
	\Delta V(S_i)=E(S_i)-E(\infty),
\end{align}
where $S_i$ denotes the distance between the baryon and meson clusters. Here, $E(S_i)$ is the adiabatic energy at the generator coordinate $S_i$, while $E(\infty)$ represents the energy when the two clusters are infinitely separated. The adiabatic energy is calculated as
\begin{align}
E(S_i)=\frac{\langle\Psi_{5q}(S_i)|H|\Psi_{5q}(S_i)\rangle}
{\langle\Psi_{5q}(S_i)|\Psi_{5q}(S_i)\rangle},
\end{align}
where $\Psi_{5q}(S_i)$ denotes the wave function of a given baryon-meson channel, and $\langle\Psi_{5q}(S_i)|H|\Psi_{5q}(S_i)\rangle$ and $\langle\Psi_{5q}(S_i)|\Psi_{5q}(S_i)\rangle$ are the corresponding diagonal matrix elements of the Hamiltonian and the normalization, respectively.

\begin{figure*}[htb]
	\centering
	\includegraphics[width=17cm]{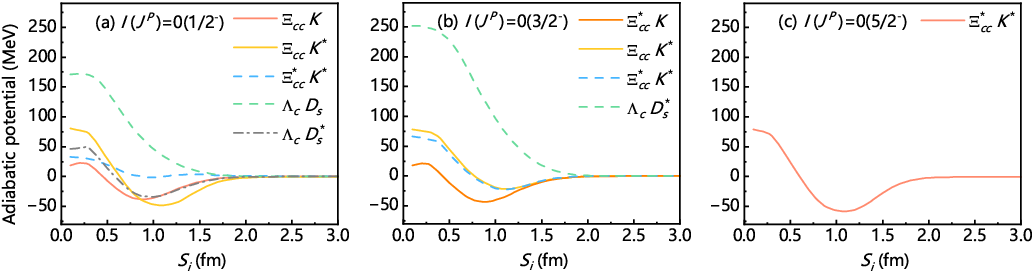}\
	\caption{\label{I=0} Adiabatic potentials for the $ccqq\bar{s}$ system with $I=0$.}
\end{figure*}

\begin{figure*}[htbp]
	\centering
	\includegraphics[width=17cm]{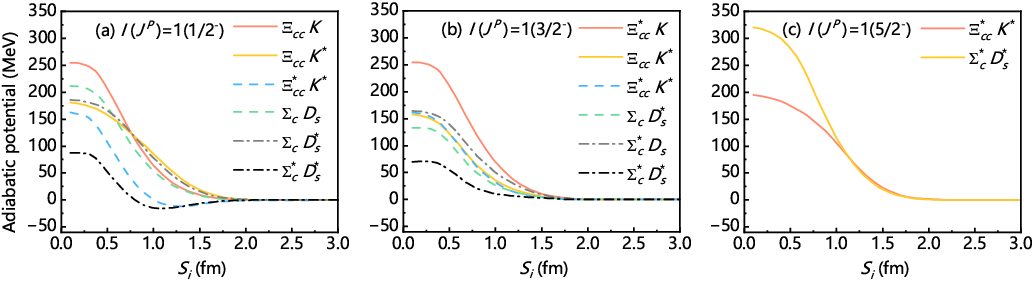}\
	\caption{\label{I=1} Adiabatic potentials for the $ccqq\bar{s}$ system with $I=1$.}
\end{figure*}

The calculated adiabatic potentials for the $ccqq\bar{s}$ system are presented in Figs.~\ref{I=0} and~\ref{I=1} for the isospin sectors $I=0$ and $I=1$, respectively. 
For the $I=0$ system, five physical channels are considered for the $J^P=1/2^-$ configuration, namely $\Xi_{cc}K$, $\Xi_{cc}K^*$, $\Xi_{cc}^{*}K^*$, $\Lambda_c D_s$, and $\Lambda_c D_s^*$. Four physical channels contribute to the $J^P=3/2^-$ configuration, including $\Xi_{cc}^{*}K$, $\Xi_{cc}K^*$, $\Xi_{cc}^{*}K^*$, and $\Lambda_c D_s^*$, while only the $\Xi_{cc}^{*}K^*$ channel is available for the $J^P=5/2^-$ configuration. 
For the $I=1$ system, six physical channels are considered for both the $J^P=1/2^-$ and $3/2^-$ configurations, whereas two channels are included for the $J^P=5/2^-$ configuration.

For the $I(J^P)=0(1/2^-)$ configuration shown in Fig.~\ref{I=0}~(a), the $\Xi_{cc}K^*$ channel exhibits the strongest attractive interaction among the five physical channels, with the adiabatic potential reaching its minimum at approximately $-48$ MeV. 
The $\Xi_{cc}K$ channel also exhibits a noticeable attractive potential well, whereas only weak attraction is found in the $\Xi_{cc}^{*}K^*$ channel.
Although the $\Lambda_c D_s^*$ channel reaches a minimum comparable to that of the $\Xi_{cc}K^*$ channel, it is also characterized by a short-range repulsive core, making its overall attraction weaker than that of the $\Xi_{cc}K^*$ channel. 
In contrast, the $\Lambda_c D_s$ channel remains predominantly repulsive throughout the interaction region.

For the $I(J^P)=0(3/2^-)$ configuration shown in Fig.~\ref{I=0}~(b), the $\Xi_{cc}^{*}K$ channel exhibits the strongest attractive interaction among the four physical channels. 
The $\Xi_{cc}K^*$ channel also possesses an attractive potential well, although it is less attractive than in the corresponding $I(J^P)=0(1/2^-)$ configuration.
In contrast, the $\Lambda_c D_s^*$ channel remains predominantly repulsive throughout the interaction region. 
The $\Xi_{cc}^{*}K^*$ channel exhibits a slightly stronger attraction than the corresponding channel in the $I(J^P)=0(1/2^-)$ configuration.
For the $I(J^P)=0(5/2^-)$ configuration shown in Fig.~\ref{I=0}~(c), only the $\Xi_{cc}^{*}K^*$ channel is involved. 
Compared with the corresponding channel in the $I(J^P)=0(1/2^-)$ and $I(J^P)=0(3/2^-)$ configurations, its attraction is significantly enhanced, leading to a much deeper and broader attractive potential well.

For the $I=1$ system shown in Fig.~\ref{I=1}, the interactions are generally dominated by repulsion. 
For the $I(J^P)=1(1/2^-)$ configuration shown in Fig.~\ref{I=1}~(a), only the $\Xi_{cc}^{*}K^*$ and $\Sigma_{c}^{*}D_s^*$ channels exhibit weak attractive pockets at intermediate distances. 
However, both channels are characterized by pronounced short-range repulsive cores, making the overall interactions predominantly repulsive. 
The remaining channels are purely repulsive over the entire interaction region. For the $I(J^P)=1(3/2^-)$ and $I(J^P)=1(5/2^-)$ configurations shown in Fig.~\ref{I=1}~(b) and (c), all considered channels remain repulsive throughout the interaction region. 
Therefore, the formation of molecular states is unlikely in the $I=1$ sector.

The adiabatic potential analysis indicates that the $I=0$ system exhibits considerably stronger attractive interactions than the $I=1$ system.
In particular, several $\Xi_{cc}^{(*)}K^{(*)}$ channels in the $I=0$ sector possess sizable attractive potential wells, whereas the $I=1$ channels are mainly dominated by repulsive interactions. 
This conclusion is consistent with the recent study in Ref~\cite{Wang:2025hhx}, where only the $I=0$ sector was further investigated because the $I=1$ interactions were found to be repulsive and unable to support bound states within the extended local hidden gauge approach.
These observations suggest that possible molecular states, if they exist, are more likely to appear in the $I=0$ sector. 
These qualitative observations will be examined quantitatively through the bound-state spectrum calculations presented in the following subsection.

\subsection{Energy spectrum}
\label{32}

To quantitatively determine whether the attractive interactions revealed by the adiabatic potentials can support molecular states, we solve the RGM equation for both single-channel and coupled-channel systems.
The calculated energies and binding energies of the $ccqq\bar{s}$ pentaquark systems are summarized in Tables~\ref{energy0} and~\ref{energy1} for the $I=0$ and $I=1$ sectors, respectively.
The first two columns list the quantum numbers and the corresponding baryon-meson channels.
The third and fourth columns give the theoretical threshold energies, $E_{\mathrm{th}}^{\mathrm{Theo}}$, and the calculated energies, $E^{\mathrm{Theo}}$, respectively.
The binding energy is defined as $E_{\mathrm{B}} = E_{\mathrm{th}}^{\mathrm{Theo}}-E^{\mathrm{Theo}}$.
Only positive values of $E_{\mathrm{B}}$ are listed in the last column, while ``UB'' denotes an unbound system.
For the coupled-channel calculations, the lowest eigenenergy together with the corresponding lowest threshold is listed in the row labeled ``Coupling''.

\begin{table}[htb]
	\caption{\label{energy0} Energies for the $ccqq\bar{s}$ pentaquark systems with $I = 0$ in single-channel and coupled-channel calculations (in MeV).}
	\begin{tabular}{l c c c c}
		\hline\hline
		~~$I(J^P)$~~ & ~~~Channel~~~ & ~~~$E_{\mathrm{th}}^{\text{Theo}}$~~~ & ~~~$E^{\text{Theo}}$~~~ & ~~~$E_{B}$~~~  \\
		\hline
		\multirow{7}{*}{$0(1/2^-)$}
		& $\Xi_{cc} K$          & 4283 & 4280 & 3   \\
		& $\Xi_{cc} K^*$        & 4680 & 4676 & 4   \\
		& $\Xi_{cc}^* K^*$      & 4705 & 4711 & UB     \\
		& $\Lambda_{c}  D_{s}$  & 4314 & 4321 & UB     \\
		& $\Lambda_{c} D_{s}^*$ & 4372 & 4371 & 1     \\ 
		& Coupling              & 4283 & 4276 & 7   \\ \hline
		
		& $\Xi_{cc}^* K$        & 4308 & 4304 & 4 \\
		& $\Xi_{cc} K^*$        & 4680 & 4685 & UB   \\
		$0(3/2^-)$
		& $\Xi_{cc}^* K^*$      & 4705 & 4709 & UB   \\
		& $\Lambda_{c} D_{s}^*$ & 4372 & 4379 & UB   \\ 
		& Coupling              & 4308 & 4290 & 18   \\ \hline
		
		$0(5/2^-)$
		& $\Xi_{cc}^* K^*$      & 4705 & 4695 & 10 \\
		\hline\hline
	\end{tabular}
\end{table}

\begin{table}[htb]
	\caption{\label{energy1} Energies for the $ccqq\bar{s}$ pentaquark systems with $I = 1$ in single-channel and coupled-channel calculations (in MeV).}
	\begin{tabular}{l c c c c}
		\hline\hline
		~~$I(J^P)$~~ & ~~~Channel~~~ & ~~~$E_{\mathrm{th}}^{\text{Theo}}$~~~ & ~~~$E^{\text{Theo}}$~~~ & ~~~$E_{\text{B}}$~~~  \\
		\hline
		
		& $\Xi_{cc} K$          & 4283 & 4290 & UB   \\
		& $\Xi_{cc} K^*$        & 4680 & 4687 & UB   \\
		& $\Xi_{cc}^* K^*$      & 4705 & 4710 & UB   \\
		$1(1/2^-)$
		& $\Sigma_{c}  D_{s}$    & 4493 & 4500 & UB     \\
		& $\Sigma_{c} D_{s}^*$   & 4551 & 4558 & UB     \\ 
		& $\Sigma_{c}^* D_{s}^*$ & 4576 & 4580 & UB   \\ 
		& Coupling               & 4283 & 4289 & UB    \\ \hline
		
		& $\Xi_{cc}^* K$        & 4308 & 4315 & UB    \\
		& $\Xi_{cc} K^*$        & 4680 & 4687 & UB   \\
		& $\Xi_{cc}^* K^*$      & 4705 & 4711 & UB   \\
		$1(3/2^-)$
		& $\Sigma_{c} D_{s}^*$   & 4551 & 4558 & UB     \\ 
		& $\Sigma_{c}^* D_{s}$   & 4518 & 4525 & UB    \\ 
		& $\Sigma_{c}^* D_{s}^*$ & 4576 & 4582 & UB  \\ 
		& Coupling               & 4308 & 4314 & UB    \\ \hline
		
		& $\Xi_{cc}^* K^*$       & 4705 & 4712 & UB   \\
		$1(5/2^-)$
		& $\Sigma_{c}^* D_{s}^*$ & 4576 & 4583 & UB \\ 
		& Coupling               & 4576 & 4583 & UB   \\
		\hline\hline
	\end{tabular}
\end{table}

Overall, the calculated energy spectra are in good agreement with the adiabatic potential analysis presented in the previous subsection.
Channels with sufficiently strong attractive interactions generally give rise to quasi-bound states, whereas those with weak attraction or predominantly repulsive interactions remain unbound.
This consistency confirms that the qualitative features inferred from the adiabatic potentials are well supported by the dynamical RGM calculations.
Furthermore, the energy spectrum provides an opportunity to examine the role of coupled-channel effects in the formation of possible molecular states.

For the $I(J^P)=0(1/2^-)$ configuration, three single-channel bound states are obtained. 
The $\Xi_{cc}K$ channel forms a bound state with a binding energy of $3$ MeV, while the $\Xi_{cc}K^*$ channel is bound with a binding energy of $4$ MeV. 
Additionally, A very weakly bound state is found in the $\Lambda_cD_s^*$ channel, with a binding energy of about only 1 MeV.
Although the $\Xi_{cc}^{*}K^*$ channel exhibits an attractive interaction, the attraction is too weak to support a bound state. 
After channel coupling is taken into account, the binding energy increases to $7$ MeV, indicating that the coupled-channel effect provides additional attraction and further stabilizes the molecular configuration.

The important role of channel coupling in the present system is also supported by the one-boson-exchange model calculation of Ref.~\cite{Sheng:2024hkf}, which also identified the coupled $\Xi_{cc} K/\Xi_{cc} K^*$ system together with the $\Xi_{cc} K^*$ channel as promising molecular candidates with $I(J^P)=0(1/2^-)$.
Since the $\Xi_{cc} K$ channel is the lowest threshold in the $I(J^P)=0(1/2^-)$ sector, the resulting bound state is stable against strong decay and therefore constitutes a promising doubly charmed molecular candidate.
Moreover, the low-lying $\Xi_{cc}K$ molecular state has also been predicted in previous investigations based on a coupled-channel unitary approach~\cite{Hofmann:2005sw}, chiral effective field theory~\cite{Guo:2017vcf}, and the extended local hidden gauge approach~\cite{Wang:2025hhx}.
As for the $\Xi_{cc} K^*$ quasi-bound state obtained in the single-channel calculation, its nature will be further clarified through the scattering analysis presented in the next subsection.

For the $I(J^P)=0(3/2^-)$ configuration, only the $\Xi_{cc}^{*}K$ channel forms a single-channel bound state with a binding energy of about $4$ MeV. 
A similar conclusion was obtained in Ref.~\cite{Wang:2025hhx}, where a virtual state was dynamically generated in the $\Xi_{cc}^{*} K$ channel close to the threshold.
Furthermore, the binding energy for the $\Xi_{cc}^{*} K$ increases remarkably to $18$ MeV in our coupled-channel calculation.
In contrast, no bound-state solutions are found for the $\Xi_{cc}K^*$, $\Xi_{cc}^{*}K^*$, and $\Lambda_cD_s^*$ channels, which can be naturally understood from the insufficient attraction revealed by the corresponding adiabatic potentials.

For the $I(J^P)=0(5/2^-)$ configuration, only the $\Xi_{cc}^{*}K^*$ channel contributes. 
Owing to its considerably stronger attraction than the corresponding channel in the $I(J^P)=0(1/2^-)$ and $0(3/2^-)$ configurations, a bound state with a binding energy of $10$ MeV is obtained. 
Since no additional coupled channels exist in the present $S$-wave calculation, channel-coupling effects are absent.
A similar $\Xi_{cc}^{*}K^*$ bound state was also predicted in Ref.~\cite{Wang:2025hhx} with a binding energy of approximately $20$ MeV. 
It should be noted that in their calculation the $J^P=1/2^-$, $3/2^-$, and $5/2^-$ states are degenerate owing to the framework employed, whereas the present quark model calculation resolves these spin configurations and predicts the bound $\Xi_{cc}^{*}K^*$ state to have the definite quantum numbers $I(J^P)=0(5/2^-)$.

The situation is entirely different for the $I=1$ system.
As shown in Table~\ref{energy1}, no bound-state solutions are found in either the single-channel or coupled-channel calculations for any of the considered quantum numbers. 
This result is consistent with the adiabatic potential analysis, where all channels are mainly dominated by repulsive interactions. 
Therefore, the present calculations indicate that the formation of $I=1$ bound states is highly unlikely within the QDCSM framework.
This conclusion is in good agreement with previous studies based on various hadronic molecular approaches~\cite{Hofmann:2005sw,Guo:2017vcf,Sheng:2024hkf,Wang:2025hhx}, all these studies predicted that the $I=1$ interactions are insufficient to support molecular bound states.

On the other hand, compact pentaquark models lead to a less consistent picture. 
For example, the CMI model~\cite{Zhou:2018bkn} predicted a compact $I(J^P)=1(3/2^-)$ pentaquark state below the $\Sigma_c^{*}D_s$ threshold, although the $\Xi_{cc}^{*}K$ channel was not included in that calculation. 
In contrast, the recent mass-splitting model~\cite{Li:2025omw}, which explicitly considered the $\Xi_{cc}^{*}K$ channel, did not predict any bound state below the $\Xi_{cc}^{*}K$ threshold. 

\subsection{Scattering phase shift}
\label{33}

To further investigate the dynamical properties of the possible molecular states, we calculate the $S$-wave scattering phase shifts for the relevant open baryon-meson channels using the KHK method. 
While the energy spectrum calculations identify states lying below the corresponding thresholds, resonance states above the lowest threshold can be reliably examined through scattering observables.
In particular, resonances are expected to manifest themselves as rapid variations of the phase shifts in the vicinity of the resonance energies.
Therefore, the scattering analysis complements the bound-state calculations and provides a more complete understanding of the spectrum of the hadronic molecular states.

\begin{figure}[htb]
	\centering
	\includegraphics[width=8cm]{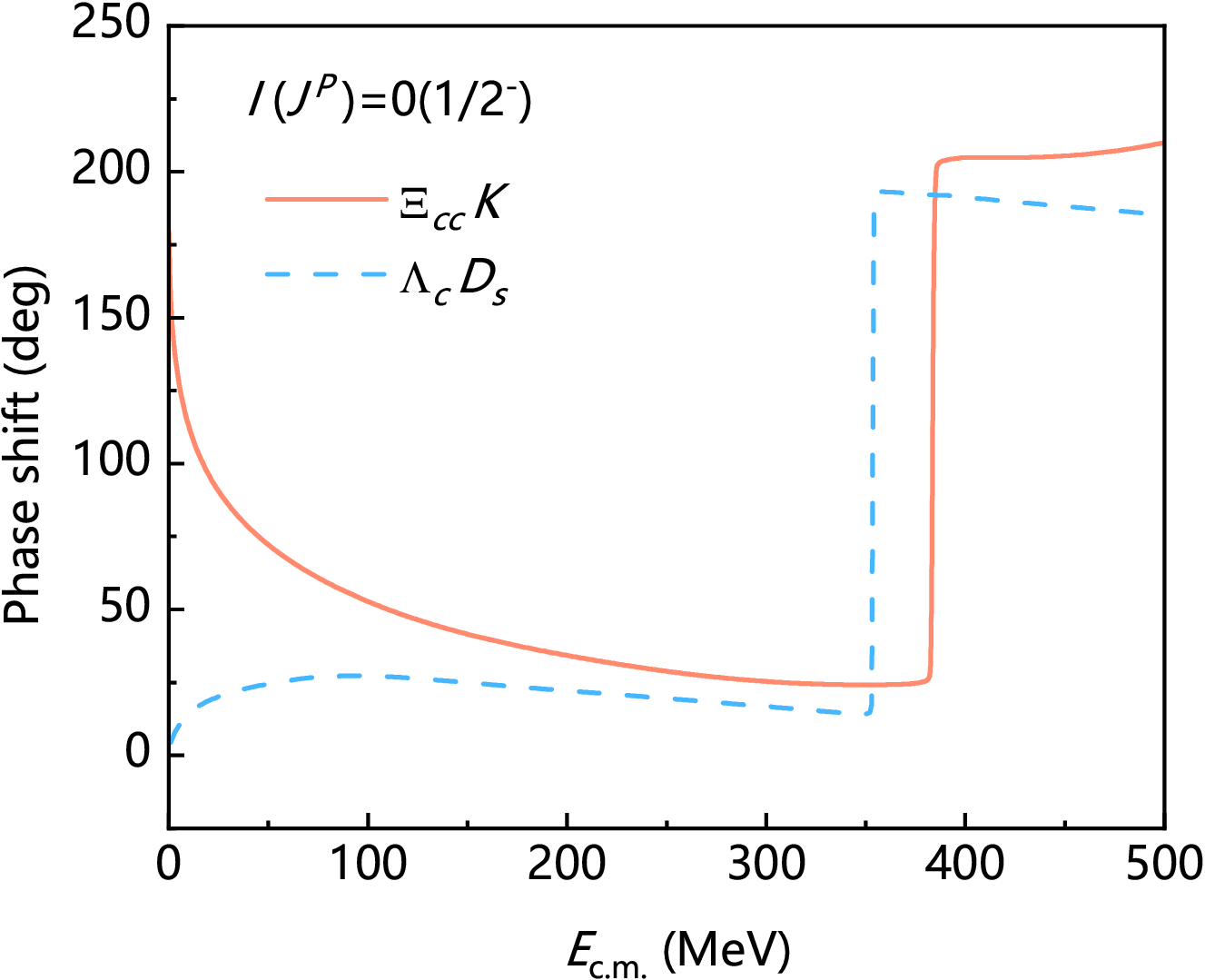}\
	\caption{Scattering phase shifts for the $\Xi_{cc}K$ and $\Lambda_cD_s$ channels with $I(J^P)=0(1/2^{-})$.}
	\label{J=0.5}
\end{figure}

To examine whether the $\Xi_{cc}K^{*}$ and $\Lambda_cD_s^{*}$ quasi-bound states obtained in the single-channel calculation can evolve into genuine resonance states after channel coupling, we calculate the $S$-wave scattering phase shifts of the $\Xi_{cc}K$ and $\Lambda_cD_s$ channels with $I(J^P)=0(1/2^{-})$, as presented in Fig.~\ref{J=0.5}.
As the center-of-mass energy approaches the $\Xi_{cc}K$ threshold, the phase shift tends to $180^\circ$, indicating the existence of a bound state below the threshold. 
This observation is consistent with the coupled-channel energy spectrum calculation, in which a  $\Xi_{cc}K$ molecular state with a binding energy of about 7 MeV is obtained.

At higher energies, both the $\Xi_{cc}K$ and $\Lambda_c D_s$ phase shifts exhibit a pronounced rapid variation, indicating that they originate from the same resonance coupled to these two open channels. 
From the phase shifts,, the resonance energy and width are extracted as:
\begin{align}
	\Xi_{cc} K^* \to \Xi_{cc} K&:~E_{r} = 4667~\text{MeV},~\Gamma= 0.06~\text{MeV}, \nonumber \\
	\Xi_{cc} K^* \to \Lambda_{c}  D_{s}&:~E_{r} = 4667~\text{MeV},~\Gamma= 0.08~\text{MeV}. \nonumber
\end{align}
The obtained resonance energy is consistent with the coupled-channel eigenenergy, demonstrating that the quasi-bound $\Xi_{cc}K^{*}$ state obtained in the single-channel calculation evolves into a genuine resonance after coupling to the open decay channels. 
Moreover, the resonance widths are only of the order of $10^{-1}$ MeV, indicating an extremely narrow resonance.

In contrast, no resonance signal associated with the weakly quasi-bound $\Lambda_cD_s^{*}$ state is observed in the calculated scattering phase shifts. 
Although a $\Lambda_cD_s^{*}$ quasi-bound state with a binding energy of only about 1 MeV is obtained in the single-channel calculation, such a shallow state is easily destabilized by channel coupling. 
The interaction with other coupled channels could raise its energy above the corresponding threshold, causing the state to dissolve into the continuum rather than survive as a genuine resonance.

\begin{figure}[htb]
	\centering
	\includegraphics[width=8cm]{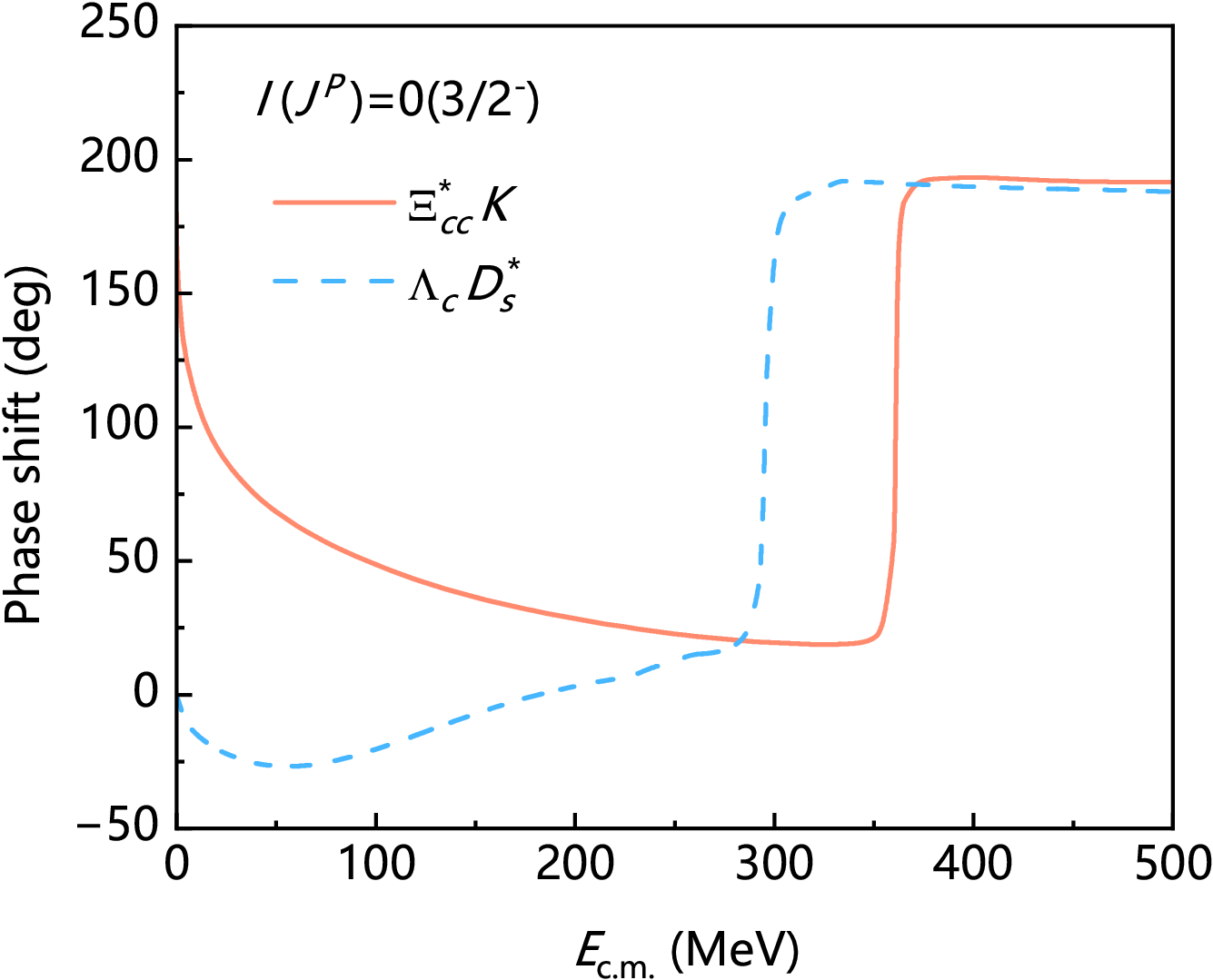}\
	\caption{Scattering phase shifts for the $\Xi^*_{cc} K$ and $\Lambda_c D^*_s$ channels with $I(J^P)=0(3/2^{-})$.}
	\label{J=1.5}
\end{figure}

For the $I(J^P)=0(3/2^{-})$ sector, we first calculate the scattering phase shifts for the lowest open channel, $\Xi^*_{cc} K$, and the results are presented in Fig.~\ref{J=1.5}.
Two notable features can be identified. 
First, the $\Xi^*_{cc} K$ phase shift starts from $180^\circ$, confirming the existence of the bound state obtained in the energy spectrum calculation with a binding energy of about 18 MeV. 
Second, a clear resonance structure is observed in the $\Xi_{cc}^{*}K$ scattering phase shift at higher energies.
This resonance is associated with the $\Xi_{cc}K^{*}$ state. 
Although the $\Xi_{cc}K^{*}$ interaction is attractive in the adiabatic potential analysis, the attraction is not sufficient to produce a bound state in the single-channel calculation. 
To clarify the origin of this resonance, we further perform a coupled-channel calculation including the $\Xi_{cc}K^{*}$ and $\Xi_{cc}^{*}K^{*}$ channels. 
It is found that the coupling to the $\Xi_{cc}^{*}K^{*}$ channel lowers the energy of the $\Xi_{cc}K^{*}$ state and generates a quasi-bound state with a binding energy of about 12 MeV. 
This quasi-bound state subsequently appears as the resonance observed in the scattering phase shifts.

The same resonance signal is also observed in the $\Lambda_{c}D_{s}^{*}$ scattering phase shift. 
According to the phase shifts, the resonance energy and width are extracted as:
\begin{align}
	\Xi_{cc} K^* \to \Xi_{cc}^* K&:~E_{r} = 4669~\text{MeV},~\Gamma = 1.7~\text{MeV}; \nonumber \\
	\Xi_{cc} K^* \to \Lambda_{c}  D_{s}^*&:~E_{r} = 4668~\text{MeV},~\Gamma = 3.8~\text{MeV}. \nonumber
\end{align}
The resonance energy agrees well with the $\Xi_{cc} K^{*}/\Xi_{cc}^{*} K^{*}$ coupled-channel eigenenergy, further confirming the resonance nature of the $\Xi_{cc}K^{*}$ state.

\begin{figure}[htb]
	\centering
	\includegraphics[width=8cm]{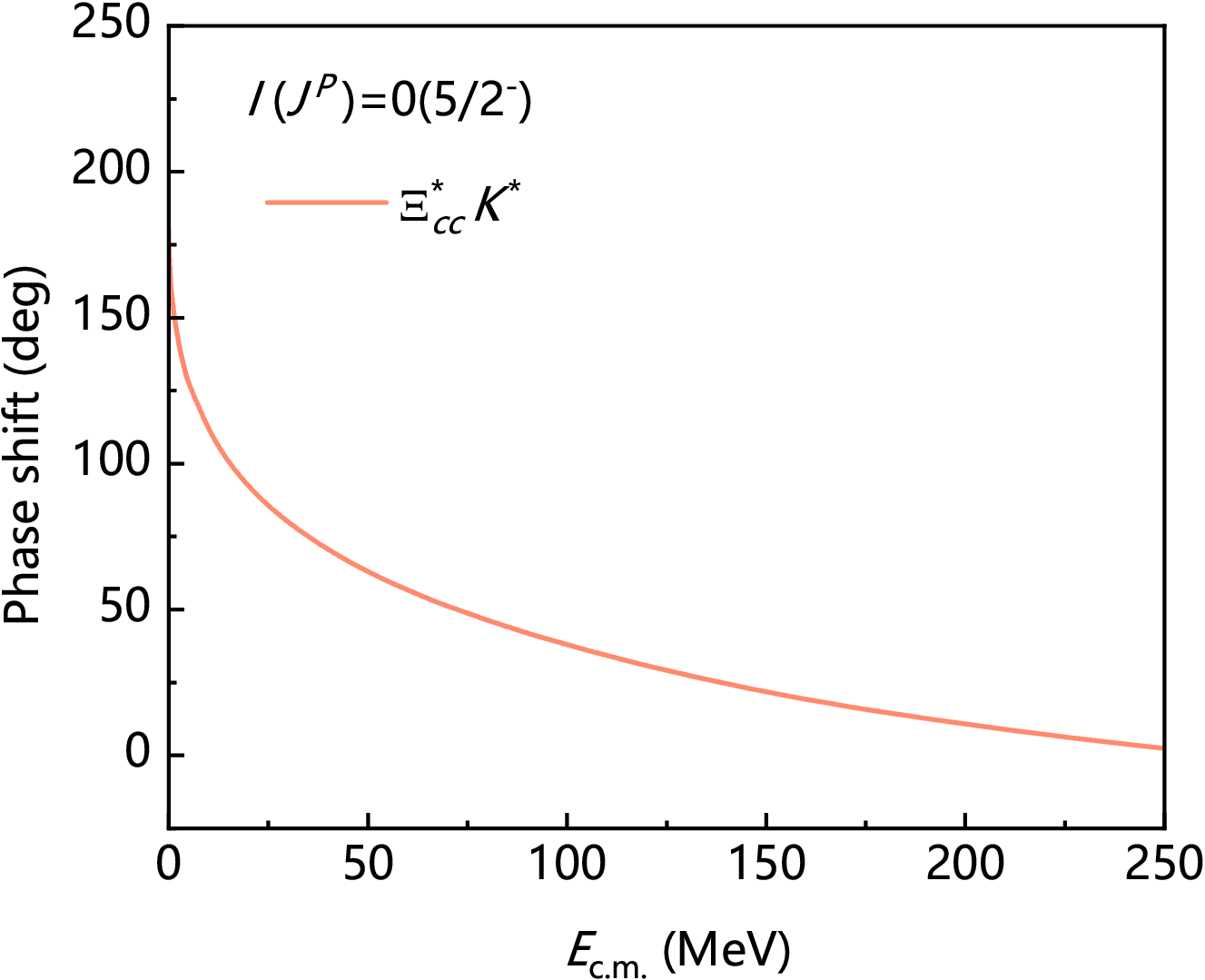}\
	\caption{Scattering phase shifts for the $\Xi^*_{cc} K^*$ channel with $I(J^P)=0(5/2^{-})$.}
	\label{J=2.5}
\end{figure}

For the $I(J^P)=0(5/2^{-})$ sector, only the $\Xi^*_{cc} K^*$ channel contributes to the scattering process. 
The corresponding phase shift is shown in Fig.~\ref{J=2.5}. 
The phase shift decreases smoothly from approximately $180^\circ$ as the center-of-mass energy increases, exhibiting the characteristic behavior of a bound state below the threshold.
The scattering phase shifts for the $I=1$ sectors with $J^P=1/2^{-}$, $3/2^{-}$, and $5/2^{-}$ have also been calculated. 
In all cases, the phase shifts exhibit neither the characteristic behavior of a bound state nor any rapid variation associated with a resonance. 
These results are consistent with the energy-spectrum calculations, which predict no bound or quasi-bound states for the corresponding channels.
Therefore, the corresponding phase shifts are not shown here for brevity.
This conclusion is also consistent with other theoretical approaches~\cite{Guo:2017vcf,Sheng:2024hkf,Wang:2025hhx}, where no bound or resonance states have been predicted.

The experimental thresholds, corrected masses, and baryon-meson distances $R$ of the molecular states obtained in the present work are summarized in Table~\ref{state}. 
Here, the masses are corrected according to:
\begin{align}
	M^{\text{Corr}}=M^{\text{Theo}}+\sum_{n} p_{n}\left[E_{\text{th}}^{\text{Exp}}(n)-E_{\text{th}}^{\text{Theo}}(n)\right],
	\label{correct}
\end{align}
in which the theoretical masses are corrected by the weighted differences between the experimental and theoretical thresholds according to the probabilities of the corresponding baryon-meson components. 
Since the experimental mass of the $\Xi^*_{cc}$ baryon has not yet been measured, we adopt the lattice QCD prediction given in Ref.~\cite{Alexandrou:2017xwd} for the mass correction.
This prediction is also consistent with the result of other theoretical approach~\cite{Ortiz-Pacheco:2023kjn}.

\begin{table}[htb]
	\caption{Thresholds, corrected masses, and baryon-meson distance $R$ of the predicted molecular states in this work. The thresholds and corrected masses are given in MeV, while the $R$ are given in fm.}
	\begin{tabular}{l c c c c}
		\hline \hline
		$I(J^P)$           & Channel            & Threshold & ~~~~Mass~~~~ & $R$  \\ \hline
		$I(J^P)=0(1/2^{-})$ & $\Xi_{cc} K$       & 4116      & 4108 & 1.81          \\
		$I(J^P)=0(1/2^{-})$ & $\Xi_{cc} K^*$     & 4513      & 4503 & 1.68   \\
		$I(J^P)=0(3/2^{-})$ & $\Xi^*_{cc} K$     & 4177      & 4168 & 1.44   \\
		$I(J^P)=0(3/2^{-})$ & $\Xi_{cc} K^*$     & 4513      & 4511 & 1.58  \\
		$I(J^P)=0(5/2^{-})$ & $\Xi^*_{cc} K^*$   & 4574      & 4564 & 1.85   \\
		\hline \hline
		\label{state}
	\end{tabular}
\end{table}

In total, five molecular states are predicted in the $I=0$ sector, including three bound states and two resonance states. 
The calculated baryon-meson distances are compatible with the expected spatial extension of hadronic molecules, further supporting their molecular interpretation.
The $\Xi_{cc}K$ state with $I(J^P)=0(1/2^{-})$ lies below the lowest strong-decay threshold and therefore can only decay through the weak interaction. 
The $\Xi_{cc}^{*}K$ and $\Xi_{cc}^{*}K^{*}$ bound states are stable against $S$-wave strong decays in the present calculation. 
Although they may still decay into the $\Xi_{cc}K$ channel through $D$-wave transitions, the corresponding decay widths are expected to be relatively small.

The two $\Xi_{cc}K^{*}$ states with $I(J^P)=0(1/2^{-})$ and $0(3/2^{-})$ appear as narrow resonances. 
Although their theoretical energies are close to each other, their corrected masses differ more significantly because the mass correction depends on the probabilities of the coupled baryon-meson components. 
In particular, $I(J^P)=0(3/2)$ state contains a $\Xi_{cc}^{*}K^{*}$ component of approximately 30\%, whereas the $I(J^P)=0(1/2^{-})$ state is predominantly composed of the $\Xi_{cc}K^{*}$ configuration.

Our results are also closely related to those obtained within the extended local hidden gauge approach~\cite{Wang:2025hhx}.
In that work, four $\Xi_{cc}^{(*)}K^{(*)}$ molecular states were predicted. Owing to the symmetry of the theoretical framework, however, the spin-parity quantum numbers of the $\Xi_{cc}^{(*)}K^{(*)}$ states other than the $\Xi_{cc}K$ state remain degenerate.
In the present coupled-channel quark model calculation, we obtain a similar spectrum while further determining the corresponding spin-parity quantum numbers and revealing the dynamical nature of these predicted doubly charmed molecular states.

Among these states, the $I(J^P)=0(1/2^{-})$ $\Xi_{cc} K$ molecular state deserves particular attention. 
Its existence has been predicted consistently in various theoretical approaches, including a coupled-channel unitary approach~\cite{Hofmann:2005sw}, chiral effective field theory~\cite{Guo:2017vcf}, constituent quark model~\cite{Sheng:2024hkf}, and the extended local hidden gauge approach~\cite{Wang:2025hhx}. The agreement among these independent frameworks strongly supports the robustness of this state, making it one of the most promising doubly charmed molecular candidates.

Moreover, the $\Xi_{cc}K$ system carries the exotic quark content $ccqq\bar{s}$. 
The absence of quark-antiquark annihilation channels also makes the $\Xi_{cc}K$ system theoretically cleaner than conventional hadrons, providing a favorable environment for studying hadronic molecular dynamics. 
Experimentally, the abundant production of doubly charmed baryons at LHCb offers a promising opportunity for future searches for such exotic molecular states. 
In addition, femtoscopic correlation measurements in high-energy collisions provide another promising approach. 
Such measurements have already proven to be a powerful tool for investigating hadron-hadron interactions and identifying hadronic molecular candidates~\cite{Hatsuda:2017uxk,Haidenbauer:2018jvl,Kamiya:2019uiw,ALICE:2020mfd,Torres-Rincon:2024znb,Jinno:2024tjh,Li:2024tvo,Etminan:2024nak,Encarnacion:2025luc,Agatao:2025ckp,Liu:2025eqw,Tang:2025bcc,Yan:2025hpa,Lin:2025mtz,Zhang:2025tfd,Zeng:2025kur,Xiong:2025bmd,STAR:2025jwe,Yan:2026yrd,Encarnacion:2026iur,Zhang:2026tqs}, and may therefore offer an alternative way to explore the $\Xi_{cc}K$ interaction. 
A detailed investigation of the $\Xi_{cc}K$ femtoscopic correlation function is planned as a natural extension of the present work.

\section{Summary}
\label{4}

In this work, we have systematically investigated the low-lying doubly charmed baryon-meson systems with the quark content $ccqq\bar{s}$ within the framework of the QDCSM. 
The possible molecular states are studied through adiabatic potential, energy-spectrum, and scattering phase shift analyses.

Five molecular candidates are obtained: the $\Xi_{cc} K$ with $I(J^P)=0(1/2^{-})$, the $\Xi_{cc} K^*$ with $I(J^P)=0(1/2^{-})$, the $\Xi^*_{cc} K$ with $I(J^P)=0(3/2^{-})$, the $\Xi_{cc} K^*$ with $I(J^P)=0(3/2^{-})$, and the $\Xi^*_{cc} K^*$ with $I(J^P)=0(5/2^{-})$. 
The scattering phase shift analysis further supports the existence of the three $S$-wave bound states, $\Xi_{cc}K$, $\Xi^*_{cc}K$, and $\Xi^*_{cc}K^*$, and demonstrates that the two $\Xi_{cc}K^{*}$ quasi-bound states evolve into genuine resonance states through channel coupling.
The calculated baryon-meson distances further support the molecular interpretation. 

These results provide useful theoretical guidance for future experimental searches for doubly charmed molecular states. 
The abundant production of charmed hadrons at LHCb provides a promising opportunity to search for such exotic doubly charmed molecular states. 
In particular, the $\Xi_{cc}K$ molecular state is of special interest, as its existence has been consistently supported by various theoretical approaches. 
Furthermore, the $\Xi_{cc}K$ interaction can be further explored through femtoscopic correlation measurements, which will be investigated in our future work.

\acknowledgments{This work is supported partly by the National Natural Science Foundation of China under Contracts Nos. 12305087 and 12575088. 	
Y.Y. is supported by the Scientific Research Foundation of Changzhou University of Information Technology under Grant No. SG210201B13003 and by the Changzhou Sci\&Tech Program under Grant No. 20260098.
Y. W is supported by the Funding for School-Level Research Projects of Yancheng Institute of Technology under Grant No. xjr2025010.
}	

\setcounter{equation}{0}
\renewcommand\theequation{A\arabic{equation}}


\begin{thebibliography}{99}

\bibitem{Klempt:2009pi} E.~Klempt and J.~M.~Richard, Baryon spectroscopy, Rev. Mod. Phys. \textbf{82}, 1095 (2010).
\bibitem{Liu:2013waa} X.~Liu, An overview of $XYZ$ new particles, Chin. Sci. Bull. \textbf{59}, 3815 (2014).
\bibitem{Richard:2016eis} J.~M.~Richard, Exotic hadrons: review and perspectives, Few Body Syst. \textbf{57}, 1185 (2016).
\bibitem{Shepherd:2016dni} M.~R.~Shepherd, J.~J.~Dudek and R.~E.~Mitchell, Searching for the rules that govern hadron construction, Nature \textbf{534}, 487 (2016).
\bibitem{Guo:2017jvc} F.~K.~Guo, C.~Hanhart, U.~G.~Mei{\ss}ner, Q.~Wang, Q.~Zhao and B.~S.~Zou, Hadronic molecules, Rev. Mod. Phys. \textbf{90}, 015004 (2018) [erratum: Rev. Mod. Phys. \textbf{94}, 029901 (2022)].
\bibitem{JPAC:2021rxu} M.~Albaladejo \textit{et al.} [JPAC], Novel approaches in hadron spectroscopy, Prog. Part. Nucl. Phys. \textbf{127}, 103981 (2022).
\bibitem{Huang:2023jec} H.~Huang, C.~Deng, X.~Liu, Y.~Tan and J.~Ping, Tetraquarks and Pentaquarks from Quark Model Perspective, Symmetry \textbf{15}, 1298 (2023).
\bibitem{Liu:2024uxn} M.~Z.~Liu, Y.~W.~Pan, Z.~W.~Liu, T.~W.~Wu, J.~X.~Lu and L.~S.~Geng, Three ways to decipher the nature of exotic hadrons: Multiplets, three-body hadronic molecules, and correlation functions, Phys. Rept. \textbf{1108}, 1 (2025).

\bibitem{Cheng:2015iom} H.~Y.~Cheng, Charmed baryons circa 2015, Front. Phys. (Beijing) \textbf{10}, 101406 (2015).
\bibitem{Lebed:2016hpi} R.~F.~Lebed, R.~E.~Mitchell and E.~S.~Swanson, Heavy-Quark QCD Exotica, Prog. Part. Nucl. Phys. \textbf{93}, 143 (2017).
\bibitem{Chen:2016qju} H.~X.~Chen, W.~Chen, X.~Liu and S.~L.~Zhu, The hidden-charm pentaquark and tetraquark states, Phys. Rept. \textbf{639}, 1 (2016).
\bibitem{Hosaka:2016pey} A.~Hosaka, T.~Iijima, K.~Miyabayashi, Y.~Sakai and S.~Yasui, Exotic hadrons with heavy flavors: $X$, $Y$, $Z$, and related states, PTEP \textbf{2016},  062C01 (2016).
\bibitem{Olsen:2017bmm} S.~L.~Olsen, T.~Skwarnicki and D.~Zieminska, Nonstandard heavy mesons and baryons: Experimental evidence, Rev. Mod. Phys. \textbf{90}, 015003 (2018).
\bibitem{Kato:2018ijx} Y.~Kato and T.~Iijima, Open charm hadron spectroscopy at $B$-factories, Prog. Part. Nucl. Phys. \textbf{105}, 61 (2019).
\bibitem{Liu:2019zoy} Y.~R.~Liu, H.~X.~Chen, W.~Chen, X.~Liu and S.~L.~Zhu, Pentaquark and Tetraquark states, Prog. Part. Nucl. Phys. \textbf{107}, 237 (2019).
\bibitem{Brambilla:2019esw} N.~Brambilla, S.~Eidelman, C.~Hanhart, A.~Nefediev, C.~P.~Shen, C.~E.~Thomas, A.~Vairo and C.~Z.~Yuan, The $XYZ$ states: experimental and theoretical status and perspectives, Phys. Rept. \textbf{873}, 1 (2020).
\bibitem{Meng:2022ozq} L.~Meng, B.~Wang, G.~J.~Wang and S.~L.~Zhu, Chiral perturbation theory for heavy hadrons and chiral effective field theory for heavy hadronic molecules, Phys. Rept. \textbf{1019}, 1 (2023).
\bibitem{Chen:2022asf} H.~X.~Chen, W.~Chen, X.~Liu, Y.~R.~Liu and S.~L.~Zhu, An updated review of the new hadron states, Rept. Prog. Phys. \textbf{86}, 026201 (2023).

\bibitem{LHCb:2017iph} R.~Aaij \textit{et al.} [LHCb], Observation of the doubly charmed baryon $\Xi_{cc}^{++}$, Phys. Rev. Lett. \textbf{119}, 112001 (2017).
\bibitem{LHCb:2018pcs} R.~Aaij \textit{et al.} [LHCb], First Observation of the Doubly Charmed Baryon Decay $\Xi_{cc}^{++}\rightarrow \Xi_{c}^{+}\pi^{+}$, Phys. Rev. Lett. \textbf{121}, 162002 (2018).

\bibitem{LHCb:2019qed} R.~Aaij \textit{et al.} [LHCb], Measurement of $\mathit{\Xi}_{cc}^{++}$ production in $pp$ collisions at $\sqrt{s}=13$ TeV, Chin. Phys. C \textbf{44}, 022001 (2020).
\bibitem{LHCb:2019epo} R.~Aaij \textit{et al.} [LHCb], Precision measurement of the $\Xi_{cc}^{++}$ mass, JHEP \textbf{02}, 049 (2020).
\bibitem{LHCb:2018zpl} R.~Aaij \textit{et al.} [LHCb], Measurement of the Lifetime of the Doubly Charmed Baryon $\Xi_{cc}^{++}$, Phys. Rev. Lett. \textbf{121}, 052002 (2018).

\bibitem{Ebert:2005xj} D.~Ebert, R.~N.~Faustov and V.~O.~Galkin, Masses of heavy baryons in the relativistic quark model, Phys. Rev. D \textbf{72}, 034026 (2005).
\bibitem{Roberts:2007ni} W.~Roberts and M.~Pervin, Heavy baryons in a quark model, Int. J. Mod. Phys. A \textbf{23}, 2817 (2008).
\bibitem{Lin:2011ti} H.~W.~Lin, Review of Baryon Spectroscopy in Lattice QCD, Chin. J. Phys. \textbf{49}, 827 (2011).
\bibitem{Can:2013tna} K.~U.~Can, G.~Erkol, B.~Isildak, M.~Oka and T.~T.~Takahashi, Electromagnetic structure of charmed baryons in Lattice QCD, JHEP \textbf{05}, 125 (2014).
\bibitem{Karliner:2014gca} M.~Karliner and J.~L.~Rosner, Baryons with two heavy quarks: Masses, production, decays, and detection, Phys. Rev. D \textbf{90}, 094007 (2014).
\bibitem{Brown:2014ena} Z.~S.~Brown, W.~Detmold, S.~Meinel and K.~Orginos, Charmed bottom baryon spectroscopy from lattice QCD, Phys. Rev. D \textbf{90}, 094507 (2014).
\bibitem{Lu:2017meb} Q.~F.~L{\"u}, K.~L.~Wang, L.~Y.~Xiao and X.~H.~Zhong, Mass spectra and radiative transitions of doubly heavy baryons in a relativized quark model, Phys. Rev. D \textbf{96}, 114006 (2017).
\bibitem{Xiao:2017udy} L.~Y.~Xiao, K.~L.~Wang, Q.~f.~Lu, X.~H.~Zhong and S.~L.~Zhu, Strong and radiative decays of the doubly charmed baryons, Phys. Rev. D \textbf{96}, 094005 (2017).

\bibitem{LHCb:2026pxn} R.~Aaij \textit{et al.} [LHCb], Observation of the Doubly Charmed Baryon {\ensuremath{\Xi}}cc+ with the LHCb Run 3 Detector, Phys. Rev. Lett. \textbf{137}, 021902 (2026).

\bibitem{BaBar:2003oey} B.~Aubert \textit{et al.} [BaBar], Observation of a narrow meson decaying to $D_s^+ \pi^0$ at a mass of 2.32-GeV/c$^2$, Phys. Rev. Lett. \textbf{90}, 242001 (2003).

\bibitem{CLEO:2003ggt} D.~Besson \textit{et al.} [CLEO], Observation of a narrow resonance of mass 2.46-GeV/$c^2$ decaying to $D^{+*}_s \pi^0$ and confirmation of the $D^*_{sJ}(2317)$ state, Phys. Rev. D \textbf{68}, 032002 (2003) [erratum: Phys. Rev. D \textbf{75}, 119908 (2007)].

\bibitem{LHCb:2020bls} R.~Aaij \textit{et al.} [LHCb], A model-independent study of resonant structure in $B^+\to D^+D^-K^+$ decays, Phys. Rev. Lett. \textbf{125}, 242001 (2020).
\bibitem{LHCb:2020pxc} R.~Aaij \textit{et al.} [LHCb], Amplitude analysis of the $B^+\to D^+D^-K^+$ decay, Phys. Rev. D \textbf{102}, 112003 (2020).

\bibitem{LHCb:2021vvq} R.~Aaij \textit{et al.} [LHCb], Observation of an exotic narrow doubly charmed tetraquark, Nature Phys. \textbf{18}, 751 (2022).

\bibitem{LHCb:2022sfr} R.~Aaij \textit{et al.} [LHCb], First Observation of a Doubly Charged Tetraquark and Its Neutral Partner, Phys. Rev. Lett. \textbf{131}, 041902 (2023).
\bibitem{LHCb:2022lzp} R.~Aaij \textit{et al.} [LHCb], Amplitude analysis of $B^0 \to \bar{D}^0 D_s^+ \pi^-$ and $B^+ \to D^- D_s^+ \pi^+$ decays, Phys. Rev. D \textbf{108}, 012017 (2023).

\bibitem{Du:2021zzh} M.~L.~Du, V.~Baru, X.~K.~Dong, A.~Filin, F.~K.~Guo, C.~Hanhart, A.~Nefediev, J.~Nieves and Q.~Wang, Coupled-channel approach to $T_{cc}^+$ including three-body effects, Phys. Rev. D \textbf{105}, 014024 (2022).
\bibitem{Albaladejo:2021vln} M.~Albaladejo, $T_{cc}^+$ coupled channel analysis and predictions, Phys. Lett. B \textbf{829}, 137052 (2022).
\bibitem{Meng:2021jnw} L.~Meng, G.~J.~Wang, B.~Wang and S.~L.~Zhu, Probing the long-range structure of the $T_{cc}^+$ with the strong and electromagnetic decays, Phys. Rev. D \textbf{104}, 051502 (2021).
\bibitem{Feijoo:2021ppq} A.~Feijoo, W.~H.~Liang and E.~Oset,$D^0 D^0 \pi^+$ mass distribution in the production of the $T_{cc}$ exotic state, Phys. Rev. D \textbf{104}, 114015 (2021).
\bibitem{Fleming:2021wmk} S.~Fleming, R.~Hodges and T.~Mehen, $T_{cc}^+$ decays: Differential spectra and two-body final states, Phys. Rev. D \textbf{104}, 116010 (2021).
\bibitem{Ling:2021bir} X.~Z.~Ling, M.~Z.~Liu, L.~S.~Geng, E.~Wang and J.~J.~Xie, Can we understand the decay width of the $T_{cc}^+$ state?, Phys. Lett. B \textbf{826}, 136897 (2022).
\bibitem{Chen:2021vhg} R.~Chen, Q.~Huang, X.~Liu and S.~L.~Zhu, Predicting another doubly charmed molecular resonance $T_{cc}'^+(3876)$, Phys. Rev. D \textbf{104}, 114042 (2021).
\bibitem{Wu:2021kbu} T.~W.~Wu, Y.~W.~Pan, M.~Z.~Liu, S.~Q.~Luo, L.~S.~Geng and X.~Liu, Discovery of the doubly charmed $T_{cc}^+$ state implies a triply charmed $H_{cccc}$ hexaquark state, Phys. Rev. D \textbf{105}, L031505 (2022).
\bibitem{Padmanath:2022cvl} M.~Padmanath and S.~Prelovsek, Signature of a Doubly Charm Tetraquark Pole in $D D^*$ Scattering on the Lattice, Phys. Rev. Lett. \textbf{129}, 032002 (2022).
\bibitem{Lyu:2023xro} Y.~Lyu, S.~Aoki, T.~Doi, T.~Hatsuda, Y.~Ikeda and J.~Meng, Doubly Charmed Tetraquark $T_{cc}^+$ from Lattice QCD near Physical Point, Phys. Rev. Lett. \textbf{131}, 161901 (2023).
\bibitem{Wang:2021mma} Q.~N.~Wang, Z.~Y.~Yang and W.~Chen, Exotic fully-heavy $Q\bar QQ\bar Q$ tetraquark states in $\mathbf{8}_{[Q\bar{Q}]}\otimes \mathbf{8}_{[Q\bar{Q}]}$ color configuration, Phys. Rev. D \textbf{104}, 114037 (2021).
\bibitem{Dong:2021bvy} X.~K.~Dong, F.~K.~Guo and B.~S.~Zou, A survey of heavy{\textendash}heavy hadronic molecules, Commun. Theor. Phys. \textbf{73}, 125201 (2021).
\bibitem{Dai:2023kwv} L.~R.~Dai, J.~Song and E.~Oset, Evolution of genuine states to molecular ones: The $T_{cc}(3875)$ case, Phys. Lett. B \textbf{846}, 138200 (2023) [erratum: Phys. Lett. B \textbf{864}, 139424 (2025)].

\bibitem{Barnes:2003dj} T.~Barnes, F.~E.~Close and H.~J.~Lipkin, Implications of a $DK$ molecule at 2.32 GeV, Phys. Rev. D \textbf{68}, 054006 (2003).
\bibitem{Kolomeitsev:2003ac} E.~E.~Kolomeitsev and M.~F.~M.~Lutz, On Heavy light meson resonances and chiral symmetry, Phys. Lett. B \textbf{582}, 39-48 (2004).
\bibitem{vanBeveren:2003kd} E.~van Beveren and G.~Rupp, Observed $D_s(2317)$ and tentative $D(2100\text{--}2300)$ as the charmed cousins of the light scalar nonet, Phys. Rev. Lett. \textbf{91}, 012003 (2003).
\bibitem{Guo:2006fu} F.~K.~Guo, P.~N.~Shen, H.~C.~Chiang, R.~G.~Ping and B.~S.~Zou, Dynamically generated 0+ heavy mesons in a heavy chiral unitary approach, Phys. Lett. B \textbf{641}, 278 (2006).
\bibitem{Gamermann:2006nm} D.~Gamermann, E.~Oset, D.~Strottman and M.~J.~Vicente Vacas, Dynamically generated open and hidden charm meson systems, Phys. Rev. D \textbf{76}, 074016 (2007).
\bibitem{Faessler:2007gv} A.~Faessler, T.~Gutsche, V.~E.~Lyubovitskij and Y.~L.~Ma, Strong and radiative decays of the $D_{s0}^*(2317)$ meson in the $DK$ molecule picture, Phys. Rev. D \textbf{76}, 014005 (2007).
\bibitem{Cleven:2014oka} M.~Cleven, H.~W.~Grie{\ss}hammer, F.~K.~Guo, C.~Hanhart and U.~G.~Mei{\ss}ner, Strong and radiative decays of the $D^*_{s0}(2317)$ and $D_{s1}(2460)$, Eur. Phys. J. A \textbf{50}, 149 (2014).
\bibitem{Mohler:2013rwa} D.~Mohler, C.~B.~Lang, L.~Leskovec, S.~Prelovsek and R.~M.~Woloshyn, $D_{s0}^*(2317)$ Meson and $D$-Meson-Kaon Scattering from Lattice QCD, Phys. Rev. Lett. \textbf{111}, 222001 (2013).
\bibitem{Lang:2014yfa} C.~B.~Lang, L.~Leskovec, D.~Mohler, S.~Prelovsek and R.~M.~Woloshyn, $D_s$ Mesons with DK and D*K Scattering Near Threshold, Phys. Rev. D \textbf{90}, 034510 (2014).
\bibitem{Ortega:2016mms} P.~G.~Ortega, J.~Segovia, D.~R.~Entem and F.~Fernandez, Molecular components in P-wave charmed-strange mesons, Phys. Rev. D \textbf{94}, 074037 (2016).
\bibitem{Du:2017ttu} M.~L.~Du, F.~K.~Guo, U.~G.~Mei{\ss}ner and D.~L.~Yao, Study of open-charm $0^+$ states in unitarized chiral effective theory with one-loop potentials, Eur. Phys. J. C \textbf{77}, 728 (2017).

\bibitem{Chen:2022svh} R.~Chen and Q.~Huang, From the isovector molecular explanation of the newly $T_{c\bar{s}}^{a0(++)}(2900)$ to possible charmed-strange molecular pentaquarks, [arXiv:2208.10196 [hep-ph]].
\bibitem{Agaev:2022eyk} S.~S.~Agaev, K.~Azizi and H.~Sundu, Modeling the resonance $T_{cs0}^{a}(2900)^{++}$ as a hadronic molecule $D^{*+}K^{*+}$, Phys. Rev. D \textbf{107}, 094019 (2023).
\bibitem{Yue:2022mnf} Z.~L.~Yue, C.~J.~Xiao and D.~Y.~Chen, Decays of the fully open flavor state $T_{c\bar{s}0}^{0}$ in a $D^{*}K^{*}$ molecule scenario, Phys. Rev. D \textbf{107}, 034018 (2023).
\bibitem{Duan:2023lcj} M.~Y.~Duan, M.~L.~Du, Z.~H.~Guo, E.~Wang and D.~Y.~Chen, Coupled-channel $D^\ast K^\ast -D_s^\ast \rho$ interactions and the origin of $T_{c\bar{s}0}(2900)$, Phys. Rev. D \textbf{108}, 074006 (2023).
\bibitem{Ke:2022ocs} H.~W.~Ke, Y.~F.~Shi, X.~H.~Liu and X.~Q.~Li, Possible molecular states of $\bar{D}^*K^* (D^*K^*)$ and new exotic states $X_0(2900)$, $X_1(2900)$, $T_{cs0}^a(2900)^0$ and $T_{cs0}^a(2900)^{++}$, Phys. Rev. D \textbf{106}, 114032 (2022).
\bibitem{Liu:2022hbk} F.~X.~Liu, R.~H.~Ni, X.~H.~Zhong and Q.~Zhao, Charmed-strange tetraquarks and their decays in the potential quark model, Phys. Rev. D \textbf{107}, 096020 (2023).
\bibitem{Yang:2023evp} X.~S.~Yang, Q.~Xin and Z.~G.~Wang, Analysis of the $T_{c\bar{s}}(2900)$ and related tetraquark states with the QCD sum rules, Int. J. Mod. Phys. A \textbf{38}, 2350056 (2023).
\bibitem{Lian:2023cgs} D.~K.~Lian, W.~Chen, H.~X.~Chen, L.~Y.~Dai and T.~G.~Steele, Strong decays of $T^a_{c{\bar{s}0}}(2900)^{++/0}$ as a fully open-flavor tetraquark state, Eur. Phys. J. C \textbf{84}, 1 (2024).
\bibitem{Wei:2022wtr} J.~Wei, Y.~H.~Wang, C.~S.~An and C.~R.~Deng, Color flux-tube nature of the states $T_{cs}(2900)$ and $T_{c\bar{s}}^a(2900)$, Phys. Rev. D \textbf{106}, 096023 (2022).
\bibitem{Ortega:2023azl} P.~G.~Ortega, D.~R.~Entem, F.~Fernandez and J.~Segovia, Novel $T_{cs}$ and $T_{c\bar{s}}$ candidates in a constituent-quark-model-based meson-meson coupled-channels calculation, Phys. Rev. D \textbf{108}, 094035 (2023).
\bibitem{Molina:2022jcd} R.~Molina and E.~Oset, $T_{c\bar{s}}(2900)$ as a threshold effect from the interaction of the $D^* K^*$, $D_s^* \rho$ channels, Phys. Rev. D \textbf{107}, 056015 (2023).
\bibitem{Duan:2023qsg} M.~Y.~Duan, E.~Wang and D.~Y.~Chen, Searching for the open flavor tetraquark $T_{c\bar{s}0}(2900)^{++}$ in the process $B^+\rightarrow K^+ D^+ D^-$, Eur. Phys. J. C \textbf{84}, 681 (2024).
\bibitem{Huang:2023fvj} Y.~Huang, H.~Hei, J.~w.~Feng, X.~Chen and R.~Wang, Production of the newly observed $\bar{T}_{c\bar{s}0}$ by kaon-induced reactions on a proton/neutron target, Phys. Rev. D \textbf{108}, 076019 (2023).
\bibitem{Wang:2023hpp} B.~Wang, K.~Chen, L.~Meng and S.~L.~Zhu, Spectrum of the molecular tetraquarks: Unraveling the $T_{cs0}(2900)$ and $\bar{T}_{c\bar{s}0}^a(2900)$, Phys. Rev. D \textbf{109}, 034027 (2024).

\bibitem{ParticleDataGroup:2026aaa} F.~Takahashi \textit{et al.} [Particle Data Group], Review of Particle Physics, Int. J. Mod. Phys. A 41, 2630011 (2026).

\bibitem{Chen:2020aos} H.~X.~Chen, W.~Chen, R.~R.~Dong and N.~Su, $X_0$(2900) and $X_1$(2900): Hadronic Molecules or Compact Tetraquarks, Chin. Phys. Lett. \textbf{37}, 101201 (2020).
\bibitem{Liu:2020nil} M.~Z.~Liu, J.~J.~Xie and L.~S.~Geng, $X_0(2866)$ as a $D^*\bar{K}^*$ molecular state, Phys. Rev. D \textbf{102}, 091502 (2020).
\bibitem{Xue:2020vtq} Y.~Xue, X.~Jin, H.~Huang and J.~Ping, Tetraquarks with open charm flavor, Phys. Rev. D \textbf{103}, 054010 (2021).
\bibitem{Hu:2020mxp} M.~W.~Hu, X.~Y.~Lao, P.~Ling and Q.~Wang, $X_0$(2900) and its heavy quark spin partners in molecular picture, Chin. Phys. C \textbf{45}, 021003 (2021).
\bibitem{He:2020btl} J.~He and D.~Y.~Chen, Molecular picture for $X_0(2900)$ and $X_1(2900)$, Chin. Phys. C \textbf{45}, 063102 (2021).
\bibitem{Agaev:2020nrc} S.~S.~Agaev, K.~Azizi and H.~Sundu, New scalar resonance $X_0(2900)$ as a molecule: mass and width, J. Phys. G \textbf{48}, 085012 (2021).
\bibitem{Wang:2021lwy} B.~Wang and S.~L.~Zhu, How to understand the $X$(2900)?, Eur. Phys. J. C \textbf{82}, 419 (2022).
\bibitem{Molina:2010tx} R.~Molina, T.~Branz and E.~Oset, A new interpretation for the $D^*_{s2}(2573)$ and the prediction of novel exotic charmed mesons, Phys. Rev. D \textbf{82}, 014010 (2010).
\bibitem{Karliner:2020vsi} M.~Karliner and J.~L.~Rosner, First exotic hadron with open heavy flavor: $cs\bar u\bar d$ tetraquark, Phys. Rev. D \textbf{102}, 094016 (2020).
\bibitem{He:2020jna} X.~G.~He, W.~Wang and R.~Zhu, Open-charm tetraquark $X_c$ and open-bottom tetraquark $X_b$, Eur. Phys. J. C \textbf{80}, 1026 (2020).
\bibitem{Wang:2020xyc} Z.~G.~Wang, Analysis of the $X_0(2900)$ as the scalar tetraquark state via the QCD sum rules, Int. J. Mod. Phys. A \textbf{35}, 2050187 (2020).
\bibitem{Zhang:2020oze} J.~R.~Zhang, Open-charm tetraquark candidate: Note on $X_0$(2900), Phys. Rev. D \textbf{103}, 054019 (2021).
\bibitem{Wang:2020prk} G.~J.~Wang, L.~Meng, L.~Y.~Xiao, M.~Oka and S.~L.~Zhu, Mass spectrum and strong decays of tetraquark ${\bar{c}}{\bar{s}} qq$ states, Eur. Phys. J. C \textbf{81}, 188 (2021).
\bibitem{Lu:2020qmp} Q.~F.~L{\"u}, D.~Y.~Chen and Y.~B.~Dong, Open charm and bottom tetraquarks in an extended relativized quark model, Phys. Rev. D \textbf{102}, 074021 (2020).
\bibitem{Tan:2020cpu} Y.~Tan and J.~Ping, $X(2900)$ in a chiral quark model, Chin. Phys. C \textbf{45}, 093104 (2021).
\bibitem{Albuquerque:2020ugi} R.~M.~Albuquerque, S.~Narison, D.~Rabetiarivony and G.~Randriamanatrika, $X_{0,1}$(2900) and $(D^-K^+)$ invariant mass from QCD Laplace sum rules at NLO, Nucl. Phys. A \textbf{1007}, 122113 (2021).
\bibitem{Liu:2020orv} X.~H.~Liu, M.~J.~Yan, H.~W.~Ke, G.~Li and J.~J.~Xie, Triangle singularity as the origin of $X_0(2900)$ and $X_1(2900)$ observed in $B^+\to D^+ D^- K^+$, Eur. Phys. J. C \textbf{80}, 1178 (2020).
\bibitem{Burns:2020epm} T.~J.~Burns and E.~S.~Swanson, Kinematical cusp and resonance interpretations of the $X(2900)$, Phys. Lett. B \textbf{813}, 136057 (2021).
\bibitem{Huang:2020ptc} Y.~Huang, J.~X.~Lu, J.~J.~Xie and L.~S.~Geng, Strong decays of ${\bar{D}}^{*}K^{*}$ molecules and the newly observed $X_{0,1}$ states, Eur. Phys. J. C \textbf{80}, 973 (2020).
\bibitem{Xiao:2020ltm} C.~J.~Xiao, D.~Y.~Chen, Y.~B.~Dong and G.~W.~Meng, Study of the decays of $S-$wave $\bar D^\ast K^\ast$ hadronic molecules: The scalar $X_0(2900)$ and its spin partners $X_{J(J=1,2)}$, Phys. Rev. D \textbf{103}, 034004 (2021).
\bibitem{Chen:2020eyu} Y.~K.~Chen, J.~J.~Han, Q.~F.~L{\"u}, J.~P.~Wang and F.~S.~Yu, Branching fractions of $B^-\rightarrow D^-X_{0,1}(2900)$ and their implications, Eur. Phys. J. C \textbf{81}, 71 (2021).
\bibitem{Chen:2021tad} H.~Chen, H.~R.~Qi and H.~Q.~Zheng, $X_1(2900)$ as a $\bar{D}_1 K$ molecule, Eur. Phys. J. C \textbf{81}, 812 (2021).
\bibitem{Lin:2022eau} Q.~Y.~Lin and X.~Y.~Wang, Searching for $ X_{0}(2900) $ and $ X_{1}(2900) $ through the kaon induced reactions, Eur. Phys. J. C \textbf{82}, 1017 (2022).
\bibitem{Dai:2022htx} L.~R.~Dai, R.~Molina and E.~Oset, Looking for the exotic $X_0(2866)$ and its $J^P=1^+$ partner in the $\bar{B}^0 \rightarrow D^{(*)+} K^{-} K^{(*) 0}$ reactions, Phys. Rev. D \textbf{105}, 096022 (2022).
\bibitem{Bayar:2022wbx} M.~Bayar and E.~Oset, Method to observe the $J^P=2^+$ partner of the $X_0(2866)$ in the $B^{+} \rightarrow D^{+} D^{-} K^{+}$ reaction, Phys. Lett. B \textbf{833}, 137364 (2022).

\bibitem{LHCb:2015yax} R.~Aaij \textit{et al.} [LHCb], Observation of $J/\psi p$ Resonances Consistent with Pentaquark States in $\Lambda_b^0 \to J/\psi K^- p$ Decays, Phys. Rev. Lett. \textbf{115}, 072001 (2015).
\bibitem{LHCb:2019kea} R.~Aaij \textit{et al.} [LHCb], Observation of a narrow pentaquark state, $P_c(4312)^+$, and of two-peak structure of the $P_c(4450)^+$, Phys. Rev. Lett. \textbf{122}, 222001 (2019).
\bibitem{LHCb:2022ogu} R.~Aaij \textit{et al.} [LHCb], Observation of a $J/\psi\Lambda$ Resonance Consistent with a Strange Pentaquark Candidate in $B^- \to J/\psi\Lambda \bar{p}$ Decays, Phys. Rev. Lett. \textbf{131}, 031901 (2023).

\bibitem{Savage:1990di} M.~J.~Savage and M.~B.~Wise, Spectrum of baryons with two heavy quarks, Phys. Lett. B \textbf{248}, 177 (1990).
\bibitem{Hu:2005gf} J.~Hu and T.~Mehen, Chiral Lagrangian with heavy quark-diquark symmetry, Phys. Rev. D \textbf{73}, 054003 (2006).
\bibitem{Guo:2013sya} F.~K.~Guo, C.~Hidalgo-Duque, J.~Nieves and M.~P.~Valderrama, Consequences of Heavy Quark Symmetries for Hadronic Molecules, Phys. Rev. D \textbf{88}, 054007 (2013).

\bibitem{Hofmann:2005sw} J.~Hofmann and M.~F.~M.~Lutz, Coupled-channel study of crypto-exotic baryons with charm, Nucl. Phys. A \textbf{763}, 90 (2005).
\bibitem{Guo:2017vcf} Z.~H.~Guo, Prediction of exotic doubly charmed baryons within chiral effective field theory, Phys. Rev. D \textbf{96}, 074004 (2017).
\bibitem{Sheng:2024hkf} L.~C.~Sheng, J.~Y.~Huo, R.~Chen, F.~L.~Wang and X.~Liu, Exploring the mass spectrum and electromagnetic property of the $\Xi_{cc} K^{(*)}$ and $\Xi_{cc} \bar{K}^{(*)}$ molecules, Phys. Rev. D \textbf{110}, 054044 (2024).
\bibitem{Wang:2025hhx} Z.~Y.~Wang and Z.~W.~Long, Prediction of $QQqq\bar{s}$ molecular pentaquarks within the extended local hidden gauge approach, Phys. Rev. D \textbf{112}, 056029 (2025).
\bibitem{Zhou:2018bkn} Q.~S.~Zhou, K.~Chen, X.~Liu, Y.~R.~Liu and S.~L.~Zhu, Surveying exotic pentaquarks with the typical $QQqq\bar{q}$ configuration, Phys. Rev. C \textbf{98}, 045204 (2018).
\bibitem{Xing:2021yid} Y.~Xing and Y.~Niu, The study of doubly charmed pentaquark $c c {\bar{q}}qq$ with the SU(3) symmetry, Eur. Phys. J. C \textbf{81}, 978 (2021).
\bibitem{Li:2025omw} S.~Y.~Li, Y.~R.~Liu, C.~R.~Shu and Z.~G.~Si, Doubly-charmed pentaquark states in a mass splitting model, Eur. Phys. J. Plus \textbf{140}, 783 (2025).
\bibitem{Rostami:2026jyz} S.~Rostami, A.~R.~Olamaei, M.~Malekhosseini and K.~Azizi, Comprehensive mass predictions: from triply heavy baryons to pentaquarks, Eur. Phys. J. C \textbf{86}, 708 (2026).

\bibitem{Wu:1996fm} G.~H.~Wu, L.~J.~Teng, J.~L.~Ping, F.~Wang and J.~T.~Goldman, Quark delocalization, color screening, and $N N$ intermediate range attraction: $P$ waves, Phys. Rev. C {\bf 53}, 1161 (1996).
\bibitem{Ping:1998si} J.~L.~Ping, F.~Wang and J.~T.~Goldman, Effective baryon baryon potentials in the quark delocalization and color screening model, Nucl. Phys. A {\bf 657}, 95 (1999).
\bibitem{Wu:1998wu} G.~h.~Wu, J.~L.~Ping, L.~j.~Teng, F.~Wang and J.~T.~Goldman, Quark delocalization, color screening model and nucleon baryon scattering, Nucl. Phys. A {\bf 673}, 279 (2000).
\bibitem{Pang:2001xx} H.~R.~Pang, J.~L.~Ping, F.~Wang and J.~T.~Goldman, Phenomenological study of hadron interaction models, Phys. Rev. C {\bf 65}, 014003 (2002).
\bibitem{Ping:2000dx} J.~L.~Ping, F.~Wang and J.~T.~Goldman, The $d^*$ dibaryon in the extended quark delocalization, color screening model, Phys. Rev. C {\bf 65}, 044003 (2002).
\bibitem{Yan:2022nxp} Y.~Yan, X.~Hu, Y.~Wu, H.~Huang, J.~Ping and Y.~Yang, Pentaquark interpretation of $\Lambda _{c}$ states in the quark model, Eur. Phys. J. C \textbf{83}, 524 (2023).
\bibitem{Yan:2023tvl} Y.~Yan, X.~Hu, H.~Huang and J.~Ping, Investigating excited $\Omega_c$ states from pentaquark perspective, Phys. Rev. D \textbf{108}, 094045 (2023).
\bibitem{Huang:2015uda} H.~Huang, C.~Deng, J.~Ping and F.~Wang, Possible pentaquarks with heavy quarks, Eur. Phys. J. C \textbf{76}, 624 (2016).
\bibitem{Huang:2018wed} H.~Huang and J.~Ping, Investigating the hidden-charm and hidden-bottom pentaquark resonances in scattering process, Phys. Rev. D \textbf{99}, 014010 (2019).

\bibitem{Wang:1992wi} F.~Wang, G.~h.~Wu, L.~j.~Teng and J.~T.~Goldman, Quark delocalization, color screening, and nuclear intermediate range attraction, Phys. Rev. Lett. \textbf{69}, 2901 (1992).
\bibitem{Chen:2011zzb} M.~Chen, H.~Huang, J.~Ping and F.~Wang, Quark model study of strange dibaryon resonances, Phys. Rev. C \textbf{83}, 015202 (2011).
\bibitem{Vijande:2004he} J.~Vijande, F.~Fernandez and A.~Valcarce, Constituent quark model study of the meson spectra, J. Phys. G \textbf{31}, 481 (2005).
\bibitem{Yan:2024usf} Y.~Yan, Q.~Huang, X.~Zhu, H.~Huang and J.~Ping, Investigating $\Xi$ resonances from a pentaquark perspective, Phys. Rev. D \textbf{110}, 014021 (2024).
\bibitem{Xu:2007oam} M.~Xu, M.~Yu and L.~Liu, Examining the crossover from hadronic to partonic phase in QCD, Phys. Rev. Lett. \textbf{100}, 092301 (2008).

\bibitem{Wheeler:1937zza} J.~A.~Wheeler, Molecular Viewpoints in Nuclear Structure, Phys. Rev. \textbf{52}, 1083 (1937).
\bibitem{Kamimura:1977okl} M.~Kamimura, Chapter V. A Coupled Channel Variational Method for Microscopic Study of Reactions between Complex Nuclei, Prog. Theor. Phys. Suppl. \textbf{62}, 236 (1977).
\bibitem{Hill:1952jb} D.~L.~Hill and J.~A.~Wheeler, Nuclear constitution and the interpretation of fission phenomena, Phys. Rev. \textbf{89}, 1102 (1953).
\bibitem{Griffin:1957zza} J.~J.~Griffin and J.~A.~Wheeler, Collective Motions in Nuclei by the Method of Generator Coordinates, Phys. Rev. \textbf{108}, 311 (1957).

\bibitem{Alexandrou:2017xwd} C.~Alexandrou and C.~Kallidonis, Low-lying baryon masses using $N_f=2$ twisted mass clover-improved fermions directly at the physical pion mass, Phys. Rev. D \textbf{96}, 034511 (2017).
\bibitem{Ortiz-Pacheco:2023kjn} E.~Ortiz-Pacheco and R.~Bijker, Masses and radiative decay widths of $S$- and $P$-wave singly, doubly, and triply heavy charm and bottom baryons, Phys. Rev. D \textbf{108}, 054014 (2023).

\bibitem{Hatsuda:2017uxk} T.~Hatsuda, K.~Morita, A.~Ohnishi and K.~Sasaki, $p\Xi^- $ Correlation in Relativistic Heavy Ion Collisions with Nucleon-Hyperon Interaction from Lattice QCD, Nucl. Phys. A \textbf{967}, 85 (2017).
\bibitem{Haidenbauer:2018jvl} J.~Haidenbauer, Coupled-channel effects in hadron-hadron correlation functions, Nucl. Phys. A \textbf{981}, 1 (2019).
\bibitem{Kamiya:2019uiw} Y.~Kamiya, T.~Hyodo, K.~Morita, A.~Ohnishi and W.~Weise, $K^-p$ Correlation Function from High-Energy Nuclear Collisions and Chiral SU(3) Dynamics, Phys. Rev. Lett. \textbf{124}, 132501 (2020).
\bibitem{ALICE:2020mfd} A.~Collaboration \textit{et al.} [ALICE], Unveiling the strong interaction among hadrons at the LHC, Nature \textbf{588}, 232 (2020) [erratum: Nature \textbf{590}, E13 (2021)].
\bibitem{Torres-Rincon:2024znb} J.~M.~Torres-Rincon, A.~Ramos and J.~Ruf{\'\i}, Two-body femtoscopy approach to the proton-deuteron correlation function, Phys. Rev. C \textbf{111}, 044906 (2025).
\bibitem{Jinno:2024tjh} A.~Jinno, Y.~Kamiya, T.~Hyodo and A.~Ohnishi, Femtoscopic study of the $\Lambda \alpha$ interaction, Phys. Rev. C \textbf{110}, 014001 (2024).
\bibitem{Li:2024tvo} H.~P.~Li, C.~W.~Xiao, W.~H.~Liang, J.~J.~Wu, E.~Wang and E.~Oset, How to unravel the nature of the $\Sigma^*(1430)(1/2-)$ state from correlation functions,, Phys. Rev. D \textbf{110}, 114018 (2024).
\bibitem{Etminan:2024nak} F.~Etminan, Exploring the $\phi-\alpha$ interaction via femtoscopic study, Phys. Lett. B \textbf{866}, 139564 (2025).
\bibitem{Encarnacion:2025luc} P.~Encarnaci{\'o}n, M.~Albaladejo, A.~Feijoo and J.~Nieves, Spectroscopic and femtoscopic insights into vector{\textendash}baryon interactions in the strangeness $-1$ sector, Eur. Phys. J. C \textbf{85}, 1347 (2025).
\bibitem{Agatao:2025ckp} B.~Agat{\~a}o, P.~Brand{\~a}o, A.~Mart{\'\i}nez Torres, K.~P.~Khemchandani, L.~M.~Abreu and E.~Oset, Correlation functions for $n\,\bar{D}_{s1}(2460)$ and $n\,\bar{D}_{s1}(2536)$, Eur. Phys. J. C \textbf{85}, 1136 (2025).
\bibitem{Liu:2025eqw} S.~W.~Liu and J.~J.~Xie, Studying the $K \Lambda$ strong interaction using femtoscopic correlation functions and the $N^*(1535)$, Phys. Rev. D \textbf{112}, 034027 (2025).
\bibitem{Tang:2025bcc} X.~M.~Tang, L.~c.~Sheng, Q.~Huang and R.~Chen, The recoil corrections, correlation functions and possible double-strange hadronic molecules, Phys. Lett. B \textbf{870}, 139896 (2025).
\bibitem{Yan:2025hpa} Y.~Yan, Q.~Huang, Q.~Wu, H.~Huang and J.~Ping, Prediction of $p\bar{\Omega}$ states and femtoscopic study, Nucl. Sci. Tech. \textbf{37}, 126 (2026).
\bibitem{Lin:2025mtz} J.~X.~Lin, P.~Encarnaci{\'o}n, M.~Albaladejo and A.~Feijoo, Signatures of odd-parity s-wave $\Xi^*$ states in femtoscopic correlation functions, Phys. Lett. B \textbf{875}, 140305 (2026).
\bibitem{Zhang:2025tfd} L.~Y.~Zhang, C.~M.~Ko, Y.~G.~Ma, Q.~Y.~Shou, K.~J.~Sun, R.~Wang and S.~Zhang, Shedding Light on (Anti-)nuclei Production with Pion-Nucleus Femtoscopy, [arXiv:2511.10298 [nucl-th]].
\bibitem{Zeng:2025kur} Z.~Zeng, B.~Chen and J.~Zhao, Relativistic corrections to hadron-hadron scattering phase shift and correlation function, Chin. Phys. C \textbf{50}, 073101 (2026).
\bibitem{Xiong:2025bmd} A.~S.~Xiong, Q.~W.~Yuan, M.~Z.~Liu, F.~S.~Yu, Z.~W.~Liu and L.~S.~Geng, Solving the inverse source problem in femtoscopy with a toy model, Chin. Phys. C \textbf{50}, 081001 (2026).
\bibitem{STAR:2025jwe} B.~E.~Aboona \textit{et al.} [STAR], First Observation of Deuteron-{\ensuremath{\Lambda}} Correlations at RHIC, Phys. Rev. Lett. \textbf{136}, 242303 (2026).
\bibitem{Yan:2026yrd} Y.~Yan, Y.~Wu, Y.~Tan, X.~Hu, Q.~Huang, H.~Huang and J.~Ping, Investigating the $\Omega \phi$ interaction and correlation functions, Phys. Rev. C \textbf{113}, 065204 (2026).
\bibitem{Encarnacion:2026iur} P.~Encarnaci{\'o}n, A.~Garc{\'\i}a-Lorenzo, M.~Albaladejo, A.~Feijoo, J.~Nieves and I.~Vida{\~n}a, Coulomb Effects in Momentum-Space Femtoscopy: A Case Study of the $\bar{K} \Omega$ System, [arXiv:2607.11321 [hep-ph]].
\bibitem{Zhang:2026tqs} K.~Zhang, X.~Wang and X.~Luo, Spin Femtoscopy: A Framework for Revealing Genuine Spin Correlations, [arXiv:2607.00413 [nucl-th]].
\end{thebibliography}
\end{document}